\documentclass[journal,12pt,onecolumn,draftclsnofoot]{IEEEtran}

\pdfoutput=1

\usepackage{amsmath}

\usepackage[cmintegrals]{newtxmath}

\usepackage{amsmath}
\usepackage{graphicx}
\usepackage{epstopdf}
\usepackage{wrapfig}
\usepackage{float}
\usepackage{cite}
\usepackage{stfloats}

\begin{document}

\title{Performance Analyses of MRT/MRC in Dual-Hop NOMA Full-Duplex AF Relay Networks with Residual Hardware Impairments}

\author{Mesut Toka, Eray G\"{u}ven, G\"{u}ne\c{s} Karabulut Kurt, O\u{g}uz Kucur

\thanks{Manuscript received Month XX, XXXX. The associate editor coordinating the review of this paper and approving it for publication was XXXX XXXX. This work was supported by the Scientific and Technological Research Council of Turkey (T\"{U}B\.{I}TAK) under Grant EEEAG 118E274.}
\thanks{M. Toka and O. Kucur are with the Department of Electronics Engineering, Gebze Technical University, Gebze/Kocaeli, 41400, Turkey (e-mail: mtoka, okucur@gtu.edu.tr). M. Toka is also with the Department of Electrical and Electronics Engineering, Ni\u{g}de \"{O}mer Halisdemir University, Ni\u{g}de, 51240, Turkey.}
\thanks{Eray G\"{u}ven and G\"{u}ne\c{s} Karabulut Kurt are with the Department of Electronics and Communications Engineering, Istanbul Technical University, Istanbul, 34469, Turkey (e-mail: guvenera, gkurt@itu.edu.tr).}}

\markboth{DRAFT~2021, To be submitted to IEEE Journal.}
{}

\maketitle

\begin{abstract}
This paper analyzes the performance of maximum-ratio transmission (MRT)/maximum-ratio combining (MRC) scheme in a dual-hop non-orthogonal multiple access (NOMA) full-duplex (FD) relay networks in the presence of residual hardware impairments (RHIs). The effects of channel estimation errors (CEEs) and imperfect successive interference cancellation are also considered for a realistic performance analysis. In the network, the base station and multiple users utilize MRT and MRC, respectively, while a dedicated relay consisting of two antennas, one for receiving and the other for broadcasting, operates in amplify-and-forward mode. For performance criterion, exact outage probability (OP) expression is derived for Nakagami-$m$ fading channels. Furthermore, a tight lower bound and asymptotic expressions are also derived to provide more insights into the obtained OP in terms of diversity order and array gain. The obtained numerical results demonstrate the importance of loop-interference cancellation process at FD relay in order for the investigated system to perform better than half-duplex-NOMA counterpart. Also, a performance trade-off between the MRT and MRC schemes is observed in the presence of CEEs among users. Furthermore, it is shown that RHIs have a significant effect on the performance of users with lower power coefficients, however it does not change the diversity order. RHIs and CEEs have the most and least deterioration effects on the system performance, respectively.       
\end{abstract}

\begin{IEEEkeywords}
Channel estimation error, full-duplex relay, imperfect successive interference cancellation, maximum-ratio combining, maximum-ratio transmission, MIMO-NOMA, residual hardware impairments.
\end{IEEEkeywords}

\IEEEpeerreviewmaketitle

\vspace{-2mm}
\section{Introduction}  
For the last decade, many researchers from academia and industry have focused on non-orthogonal multiple access (NOMA) in order to overcome challenges caused by smart devices with massive connectivity and fulfill requirements of forthcoming generations wireless networks since the existing orthogonal multiple access (OMA) techniques are limited in terms of spectral efficiency and massive connectivity \cite{QCLi,LiuY,Aldababsa}. The most important key feature of NOMA is to serve multiple users in the same resources (time/frequency/code) by allocating different power coefficients, thus fairness among users can be ensured. On the other hand, successive interference cancellation (SIC) technique is applied by users to separate the superposed signals and obtain the desired information related to user \cite{Saito1}. So far, NOMA, especially power-domain NOMA, has been widely investigated in the literature. In \cite{DingZ}, a downlink NOMA network based on cell-clustering, where distances between the base station (BS) and users are subjected to uniform distribution, is investigated. In \cite{Timotheou}, the authors have focused on optimization of power coefficients in order to ensure maximizing the fairness among user. In \cite{Gui}, a new definition to measure the fairness, which evaluates rate of each user by accounting for the fraction of total power allocated to it, for NOMA networks is proposed. On the other hand, in order to exploit benefits of spatial diversity, multiple-input multiple-output (MIMO) techniques are also considered in NOMA networks \cite{DingZ1}. Accordingly, in \cite{QiSun}, ergodic capacity maximization problem, where cost function is subjected to total transmit power constraint and minimum rate constraint related to weak user, is investigated for a two-user MIMO-NOMA network. In \cite{MZeng}, the authors have analyzed sum and ergodic capacity of MIMO-NOMA system, in which multiple users are grouped into a cluster according to certain algorithms, and demonstrated that the sum capacity is inversely proportional to the number of users within cluster. In addition, several beamforming techniques have been investigated for multiple-input single-output (MISO) and/or MIMO NOMA networks in order to either maximize capacity and fairness or minimize transmit power \cite{QZhang,DingZ2,XChen,FAlavi}. In \cite{MtokaCL}, the authors have analyzed the outage probability (OP) of Alamouti space-time block coding (STBC) \cite{Tarokh} in MISO multi-user NOMA network over independent and identically distributed (i.i.d.) Nakagami-$m$ fading channels. In \cite{Mtoka}, the same network considered in \cite{MtokaCL} has been generalized to all orthogonal STBC (OSTBC) codes. The authors have analyzed the OP and ergodic capacity over i.i.d. Nakagami-$m$ fading channels by also considering the effects of channel estimation errors (CEEs), feedback delay (FBD) and imperfect SIC (ipSIC). Although using multiple antennas at the transmitter and receiver ends provides improved transmission reliability and spatial diversity, wireless systems suffer from hardware complexity and high power consumption. Therefore, the authors of \cite{YYu1} proposed several computationally efficient antenna selection algorithms for two-user MIMO-NOMA scenarios based on fixed power and cognitive radio-inspired power allocations in order to maximize the system sum-rate. In \cite{AldababsaMajTAS}, a novel antenna selection scheme for single-hop MIMO-NOMA based on decision of majority of users differing from \cite{YYu1} has been proposed and OP performance has been investigated over Nakagami-$m$ fading by also considering effects of CEEs and FBD.

Moreover, since relaying techniques offer extending coverage area and establish reliable communication under heavy channel environments and huge obstacles, recently, NOMA has been extended to cooperative transmission \cite{DingZ3,JBKim}. In \cite{JJmen}, the OP analysis of a dual-hop amplify-and-forward (AF) relaying NOMA network, where the BS and users are equipped with single antenna, has been conducted over Nakagami-$m$ fading under effect of CEEs. In \cite{ZhangY}, the authors have investigated transmit antenna selection (TAS) at the BS and maximal-ratio combining (MRC) at users in a dual-hop AF relaying NOMA system over Nakagami-$m$ fading by considering CEEs. In \cite{XYan}, OP and ergodic capacity of cooperative two-user NOMA networks with direct link, where single relay among multiple relays and single antenna at users are selected, has been analyzed in the presence of CEEs and ipSIC. The authors of \cite{HLi} have analyzed joint relay and antenna selection problem in cooperative two-user NOMA network based on coordinated direct and relay transmission structure in terms of OP over Rayleigh fading channels. In \cite{AldababsaMRTRAS}, the authors have analyzed OP of a dual-hop MIMO-NOMA network, where maximal-ratio transmission (MRT)/receive antenna selection (RAS) are adopted in both hops, over Nakagami-$m$ fading channels in the presence of CEEs. Although reliability and performance of NOMA networks have been increased by aforementioned cooperative studies, however they are all based on half-duplex (HD) relaying which has limited spectral efficiency because of allocation of two orthogonal channels for transmission. On the other hand, full-duplex (FD) relaying technique has been regarded as a promising solution since the reception and transmission can be realized at the same time/frequency yielding double capacity \cite{Duarte}. However, FD relay has a major drawback named as loop-interference (LI) caused by signal leakage between the transmitter and receiver antennas. Fortunately, thanks to the advances on antenna technologies and signal processing approaches, effect of LI can be reduced to a sufficient level in order for FD relaying be feasible in a practical manner \cite{Duarte,Rodriguez}. Therefore, the authors of \cite{ZhongC} and \cite{TMCC} have considered using FD relay in NOMA systems in order to overcome spectral efficiency loss caused by HD relays. Also, in \cite{YAlsaba}, a cooperative two-user NOMA network, where the BS adopts zero-forcing (ZF) beamforming and the strong user acts as a decode-and-forward (DF) FD relay to assist weak user, has been proposed. The authors have analyzed OP of the network over Rayleigh fading channels by also taking into account of energy harvesting at the relay. In \cite{Mohammadi}, a cooperative two-user NOMA network based on cognitive radio has been considered and joint beamforming optimization problem for transmit/receive ends at DF FD relay has been evaluated. The authors have also investigated MRT/ZF, ZF/MRC and ZF/ZF schemes at the relay in a comparative manner. 

In practice, radio-frequency hardware components at the transmitter and receiver suffer from different impairments caused by high power amplifier non-linearity, in-phase/quadrature (I/Q) imbalance and phase noise, which seriously deteriorate the system performance due to mismatch between the desired and actual signal \cite{Schenk}. In order to compensate the influence of these hardware impairments (HIs), a lot of efforts in developing appropriate approaches have been made in the literature, however, there still exist residual HIs (RHIs) which can not be overlooked in real-life deployments \cite{Bjornson,Bjornson2}. Although some of the aforementioned studies on NOMA networks consider only CEEs, FBD and ipSIC without RHIs in a practical manner, nevertheless, there are some works paying attention to effects of RHIs in NOMA networks. Particularly, in \cite{XLi3}, the impact of I/Q imbalance impairment in a secure single-input multiple-output (SIMO)-NOMA network consisting of one BS, multiple legitimate users and an eavesdropper has been investigated over Rayleigh fading channels. In order to exploit receive diversity, RAS scheme has been applied at receivers of all users and eavesdropper. In \cite{MtokaMRTRAS}, outage performance of hybrid MRT/RAS scheme is investigated in single-hop multi-user NOMA system in the presence of RHIs together with CEEs and ipSIC. In \cite{FDing}, impact of RHIs on dual-hop multi-user NOMA network with AF relay has been investigated over Nakagami-$m$ fading channels in terms of OP and ergodic rate. In \cite{XLi}, the authors analyzed OP and ergodic capacity of single-hop and dual-hop NOMA AF relay networks over $\alpha-\mu$ fading channels in the presence of CEEs, ipSIC and RHIs. In \cite{XLi2}, effects of RHIs together with CEEs and ipSIC have been analyzed in a dual-hop NOMA AF relaying network, where multiple users are separated into multiple clusters and the BS communicates all clusters according to OMA while adopts NOMA scheme to serve users in a cluster, over Nakagami-$m$ fading channels. The authors have also considered energy harvesting at the relay, and obtained exact OP and ergodic capacity expressions for two users in a cluster. In \cite{CBLe}, a cooperative two-user NOMA network, where all nodes are equipped with single antenna and a DF hybrid HD/FD relay assists the communication between the BS and far user, has been investigated over Rayleigh fading channels in the presence of RHIs. In order to demonstrate the level of system performance, OP and ergodic capacity expressions have been obtained. In \cite{CDeng}, the authors have considered the same system of \cite{CBLe} without direct link between the relay and near user, and analyzed OP and ergodic rate over Rician fading channels.                                           

As mentioned above, there are many studies investigate effects of RHIs on single-hop and/or dual-hop NOMA networks with or without CEEs/ipSIC in the literature. However, the majority of them consider systems as consisting of HD relays and nodes equipped with single antenna except studies of \cite{XLi3,CBLe,CDeng}. In \cite{XLi3}, users and an eavesdropper are equipped with multiple antennas to adopt RAS scheme while FD relaying is considered in \cite{CBLe} and \cite{CDeng}. Therefore, investigations on the impact of RHIs on both MIMO-NOMA and FD relaying based cooperative NOMA, to the best of our knowledge, are still limited. Motivated by \cite{XLi3,CBLe,CDeng}, in this paper, we investigate a dual-hop multi-user AF FD relaying based MIMO-NOMA system, where MRT and MRC schemes are exploited at the BS and users, respectively, over i.i.d. Nakagami-$m$ fading channels in the presence of RHIs. The key contributions of the paper are summarized as follows:
\begin{itemize}
	\item Unlike the existing studies on cooperative NOMA with RHIs, we consider using multiple antennas at the BS and users while the relay operates in FD mode. Furthermore, our analyses have been conducted for a generic channel model, Nakagami-$m$ fading, and LI link at the relay is also assumed to exposed to fading variations.
	\item In order to provide a realistic analysis, CEEs and ipSIC have been taken into account. To characterize the system performance, exact OP expression for any user has been derived. Moreover, a tight lower bound and simple asymptotic expressions have been obtained to provide further insights, such as diversity behavior of the system. The investigated network has been compared to HD-NOMA and FD-OMA counterparts.
	\item We have demonstrated that the quality of LI cancellation process is quite crucial for the investigated system to outperform HD-NOMA counterpart, even error floors may exist under the worst case of cancellation process. In terms of users with lower power allocations, MRT performs better than MRC without CEEs, while there is a trade-off between both schemes in the presence of CEEs. In addition, RHIs have much more effect on the performance of users with lower power allocations while do not change the diversity order of all users. Moreover, imperfections which have the most and least deterioration effects on the performance are RHIs and CEEs, respectively.            
\end{itemize}     
       
\subsection{Organization and Notations}  
The rest of the paper is given as follows. In Section II, we introduce the system model and channel statistics including practical imperfections in detail. In Section III, we derive the exact OP expression for any user together with lower bound and asymptotic approximations. Numerical results including comparisons are illustrated in Section IV. Finally, conclusions are interpreted in Section V.

\textit{Notation:} Bold lowercase letter and $\|\cdot\|$ denote vectors and Euclidean norm while $(\cdot)^H$ indicates Hermitian transpose of a vector. $\mathbb{CN}(0,\sigma^2)$ is used to represent the complex Gaussian distribution with zero mean and variance of $\sigma^2$. While $Pr(\cdot)$ denote the probability of an event, $E\left[\cdot\right]$ represents the expectation operator. $f_X(\cdot)$ and $F_X(\cdot)$ indicate the probability density function (PDF) and cumulative distribution function (CDF) of a random variable $X$, respectively.

\section{System Model}
We consider a dual-hop power domain downlink NOMA network, where one BS ($S$) communicates with $L$ users ($U_l, l=1,2,\cdots,L$) with assistance of an FD AF relay ($R$). We assume that the direct link is not available due to huge obstacles and heavy environment conditions. The BS equipped with $N_S$ antennas transmits information by applying MRT beamforming technique while users with $N_D$ antennas combine received signals according to MRC scheme. Note that MRT and MRC schemes are the optimum ones among transmit and receive diversity techniques, respectively \cite{LoT,Simon}. On the other hand, the FD relay has two antennas, one for receiving and the other for broadcasting. $\mathbf{h}_{SR}=\left\lbrace h_{SR}^i \right\rbrace_{1\times N_S}$ ($1\leq i\leq N_S$), $\mathbf{h}_{l}=\{h_{l}^j\}_{N_D\times 1} $ ($1\leq j\leq N_D$) and $h_{LI}$ denote channel coefficients corresponding to $S-R$, $R-U_l$ and $R-R$ links. Since channels are assumed to be distributed as i.i.d. Nakagami-$m$, squared of channel gains follow Gamma distribution, thus powers of links can be obtained by $\Omega_{SR}=E\left[ |h^i_{SR}|^2\right] =d_{SR}^{-\alpha}$, $\Omega_{l}=E[ |h^j_{l}|^2] =d_{l}^{-\alpha}$ and $\Omega_{LI}=E[ |h_{LI}|^2] =\lambda P_R^{\mu-1}$, respectively. $d_{SR}$ and $d_{l}$ denote the normalized distances of $S-R$ and $R-U_l$ links, respectively, $\alpha$ is the path loss exponent. $\lambda$ ($\lambda>0$) and $\mu$ ($0\leq\mu\leq1$) represent the quality of LI cancellation process at $R-R$ link\footnote{It is worthwhile noting that the FD relay suffers from a LI effect between transmit and receive antennas due to its inherent simultaneous transmission at the same time/frequency. Although LI effects have been mitigated somehow, there still remain some residual LI effects. Without loss of generality, we consider the residual LI model determined according to active and/or passive interference cancellation as in \cite{Duarte} and \cite{Rodriguez}.}. Since effects of CEEs are also considered to be more practical, by following linear minimum mean square error estimation method, the BS and relay estimates channel coefficients in the training period, thus erroneously estimated channel coefficient vectors of $S-R$ and $R-U_l$ links can be represented by $\mathbf{h}_{SR}=\hat{\mathbf{h}}_{SR}+\boldsymbol{\varepsilon}_{e,SR}$ and $\mathbf{h}_{l}=\hat{\mathbf{h}}_{l}+\boldsymbol{\varepsilon}_{e,l}$, respectively\footnote{Note that the utilized channel estimation model is widely considered in the existing literature \cite{Mtoka,AldababsaMRTRAS,XLi}.}. $\boldsymbol{\varepsilon}_{e,SR}$ and $\boldsymbol{\varepsilon}_{e,l}$ are error vectors resulting from imperfect estimation process and can be modeled as $\boldsymbol{\varepsilon}_{e,SR}\sim\mathbb{CN}(0,\sigma_{e,SR}^2)$ and $\boldsymbol{\varepsilon}_{e,l}\sim\mathbb{CN}(0,\sigma_{e,l}^2)$ with variances of $\sigma_{e,SR}^2=\Omega_{SR}-\hat{\Omega}_{SR}$ and $\sigma_{e,l}^2=\Omega_{l}-\hat{\Omega}_{l}$, respectively \cite{Medard}.                     

Since the investigated system is based on NOMA transmission, the BS transmits superimposed signals represented by $x(n)=\sum_{i=1}^{L}\sqrt{P_Sa_i}s_i(n)$ in the $n$th time interval, where $P_S$ and $a_i$ represent transmit power at the BS and power allocation coefficient intended to $i$th user ($\sum_{i=1}^{L}a_i=1$), respectively. Note that we represented signals according to time index due to the FD relay transmission. Then, the received signal at the relay under RHIs effect can be represented as
\begin{equation}\label{eq:1}
	y_R(n)=\mathbf{h}_{SR}(\mathbf{w}(n)x(n)+\eta_{SR}(n))+h_{LI}s_R(n)+n_R(n),
\end{equation}
where $n_R(n)\sim\mathbb{CN}(0,\sigma_{R}^2)$ is Gaussian noise at $R$ and $\eta_{SR}(n)\sim\mathbb{CN}(0,\kappa_{SR}^2P_S)$ represents aggregate distortion noise resulting from RHIs at $S-R$ link. $\kappa_{SR}^2=(\kappa_{S}^t)^2+(\kappa_{R}^r)^2$ denotes aggregate power level of RHIs, where $\kappa_{S}^t$ and $\kappa_{R}^r$ are impairment levels \cite{Schenk,Bjornson}. Since FD relay applies LI cancellation methods, we assume that the impact of RHIs distortion noise at link $R-R$ is absorbed by the LI cancellation parameter as in \cite{Nguyen}. $\mathbf{w}(n)=\hat{\mathbf{h}}_{SR}^H/\|\hat{\mathbf{h}}_{SR}\|$ is MRT weight vector at the BS, subjected $\|\mathbf{w}(n)\|^2=1$. Also, $s_R(n)=Gy_R(n-\tau)$ denotes the signal to be transmitted from the relay, where $\tau$ is processing delay of the FD transmission and $G$ is the amplification factor which can be obtained as
\begin{equation}\label{eq:2}
	G=\sqrt{\frac{P_R}{P_S(\|\hat{\mathbf{h}}_{SR}\|^2+\sigma_{e,SR}^2)(1+\kappa_{SR}^2)+P_R|h_{LI}|^2+\sigma_{R}^2}}.
\end{equation}

Afterwards, the received signal vector at the $l$th user can be written as $\mathbf{y}_{U_l}(n)=\mathbf{h}_{l}(s_R(n)+\eta_{RU}(n))+\mathbf{n}_l(n)$, where $\mathbf{n}_l(n)=\{n_l^j(n)\}_{N_D\times 1}$ is Gaussian noise vector (whose entries are subjected to $\mathbb{CN}(0,\sigma_l^2)$) at $l$th user. Also, $\eta_{RU}(n)\sim\mathbb{CN}(0,\kappa_{RU}^2P_R)$ denotes the aggregate distortion noise of RHIs at $R-U_l$ link, where $\kappa_{RU}^2=(\kappa_{R}^t)^2+(\kappa_{U}^r)^2$ is aggregate power level of RHIs. Without loss of generality, as in \cite{FDing}, we consider that users have the same effect of hardware impairments; such that $\kappa_{U_l}^r\stackrel{\triangle}{=}\kappa_{U}^r$. Then, if we substitute $s_R(n)$ into $\mathbf{y}_{U_l}(n)$, the received signal vector at $l$th user can be rewritten with the help of (\ref{eq:1}) as 
\begin{equation}\label{eq:3}
	\begin{split}
		\mathbf{y}_{U_l}(n)&=\mathbf{h}_{l}G\mathbf{h}_{SR}\bigg[\underbrace{\mathbf{w}(n-\tau)\sqrt{P_Sa_l}s_l(n-\tau)}_{\textit{desired signal}} +\underbrace{\mathbf{w}(n-\tau)\sum\nolimits_{p=1}^{l-1}\sqrt{P_Sa_p}s_p(n-\tau)}_{\textit{ipSIC term}} \\
		& +\underbrace{\mathbf{w}(n-\tau)\sum\nolimits_{k=l+1}^{L}\sqrt{P_Sa_k}s_k(n-\tau)}_{\textit{IUI term}}+\underbrace{\eta_{SR}(n-\tau)}_{\textit{RHIs at S-R}}\bigg] \\
		&+\mathbf{h}_{l}\bigg[ G\bigg( \underbrace{h_{LI}s_R(n-\tau)}_{\textit{LI term}}+n_R(n-\tau)\bigg) + \underbrace{\eta_{RU}(n)}_{\textit{RHIs at $R-U_l$}}\bigg]+ \mathbf{n}_l(n).	
	\end{split}
\end{equation} 

The received signals by $N_D$ antennas at the $l$th user are combined according to MRC technique as $y_{U_l}^{MRC}=\mathbf{w}_{\text{MRC}}(n)\mathbf{y}_{U_l}(n)$, where $\mathbf{w}_{\text{MRC}}(n)=\hat{\mathbf{h}}_{l}^H/\|\hat{\mathbf{h}}_{l}\|$ is MRC weight vector subject to $\|\mathbf{w}_{\text{MRC}}(n)\|^2=1$.

\section{Performance Analyses}
In this section, end-to-end ($e2e$) signal-to-interference-distortion plus noise ratio (SIDNR) expression is derived. Then, the exact OP for any user is obtained together with lower bound and asymptotic expressions to provide further insights into the system performance.

\subsection{Derivation of $e2e$ SIDNR}
According to NOMA transmission, weaker users (with poorer channel qualities) are allocated higher power levels at the BS for ensuring the fairness. Therefore, in the training period, the relay estimates effective channel gains of $R-U_l$ links by using pilot symbols sent from all users such that they are ordered as $\|\hat{\mathbf{h}}_{RU_1}\|^2\leq\|\hat{\mathbf{h}}_{RU_2}\|^2$ $\cdots\leq\|\hat{\mathbf{h}}_{RU_L}\|^2$ without loss of generality, and then transmits the ordering to the BS and users at the same time. Thus, the BS allocates power coefficients to users as $a_1>a_2>\cdots>a_L$ by using the ordering. Also, the BS estimates channel gains of $S-R$ link to apply MRT beamforming. Since SIC is carried out at users, any stronger user $l$ detects and removes signal of the weaker user $j$, where $j<l$. On the other hand, signal of the stronger user $k$ is considered as interference noise by user $l$, where $k>l$ and also named as inter-user interference (IUI). Consequently, by using (\ref{eq:2}) and (\ref{eq:3}), instantaneous SIDNR defined as the $l$th user erroneously decodes the signal of $j$th user ($j\leq l$) is given as
\begin{equation}\label{eq:4}
	\gamma_{U_{j\rightarrow l}}=\dfrac{\psi_1\psi_2\bar{\gamma}^2a_j}{\psi_1\psi_2\bar{\gamma}^2(\xi_j+\tilde{\xi}_j+\vartheta_1)+\psi_1\bar{\gamma}\vartheta_2\vartheta_3+(\psi_2\bar{\gamma}+\vartheta_2)(\psi_3\bar{\gamma}\vartheta_4+\vartheta_5)\vartheta_3}
\end{equation} 

In (\ref{eq:4}), $\bar{\gamma}=P/\sigma^2$ represents average signal-to-noise ratio (SNR), where $P_S=P_R=P$ is assumed for mathematical simplicity, while $\xi_j=\sum_{k=j+1}^{L}a_k$ and $\tilde{\xi}_j=\sum_{p=1}^{j-1}a_p\sigma_{ipsic}^2$ are IUI and ipSIC terms, respectively. Without loss of generality, we assume that ipSIC is subject to Gaussian distribution with power $\sigma_{ipsic}^2$ ($0\leq\sigma_{ipsic}^2\leq1$) as in \cite{XLi,HasanA,Tweed}. Also, $\psi_1\stackrel{\triangle}{=}\|\hat{\mathbf{h}}_{SR}\|^2$, $\psi_2\stackrel{\triangle}{=}\|\hat{\mathbf{h}}_{l}\|^2$ and $\psi_3\stackrel{\triangle}{=}|h_{LI}|^2$ definitions are made to simplify analyses. In addition, constant variables $\vartheta$ are given below:
\begin{equation}\label{eq:5}
\begin{split}
\vartheta_1&\stackrel{\triangle}{=}\kappa_{SR}^2+\kappa_{RU}^2(1+\kappa_{SR}^2)~,~~\vartheta_2\stackrel{\triangle}{=}\bar{\gamma}\sigma_{e,l}^2+\frac{1}{1+\kappa_{RU}^2} \\
\vartheta_3&\stackrel{\triangle}{=}(1+\kappa_{RU}^2)(1+\kappa_{SR}^2)~~,~~\vartheta_4\stackrel{\triangle}{=}\frac{1}{1+\kappa_{SR}^2}  \\
\vartheta_5&\stackrel{\triangle}{=}\bar{\gamma}\sigma_{e,SR}^2+\frac{1}{1+\kappa_{SR}^2}. 
\end{split}
\end{equation}

\subsection{Outage Probability Analysis}
The outage event for the $l$th user can be defined as the $l$th user can not decode its own signal or the $j$th user's signal ($1\leq j\leq l$). Thus, let us define $E_{l,j}=\{\gamma_{U_{j\rightarrow l}}>\gamma_{th,j}\}$ as the event that the $l$th user can decode $j$th user's signal, where $\gamma_{th,j}=2^{R_0}-1$ ($R_0$: bits per channel in use (BPCU)) is the target threshold SIDNR for FD transmission. With the help of (\ref{eq:4}), the event of $E_{l,j}$ can be expressed as 
\begin{equation}\label{eq:6}
	\begin{split}
		E_{l,j}&=\left\lbrace \left(\psi_2-\vartheta_2\vartheta_3\delta_j \right)\psi_1>\frac{(\psi_2\bar{\gamma}+\vartheta_2)(\psi_3\bar{\gamma}\vartheta_4+\vartheta_5)\vartheta_3\delta_j}{\bar{\gamma}}\right\rbrace  \\
		&=\left\lbrace \psi_1>\frac{(\psi_2\bar{\gamma}+\vartheta_2)(\psi_3\bar{\gamma}\vartheta_4+\vartheta_5)\vartheta_3\delta_j}{\bar{\gamma}(\psi_2-\vartheta_2\vartheta_3\delta_j)},\psi_2>\vartheta_2\vartheta_3\delta_j\right\rbrace,  
	\end{split}
\end{equation}
where the second equality is obtained subjected to the condition of $a_j-\gamma_{th,j}(\xi_j+\tilde{\xi}_j+\vartheta_1)>0$ and also $\delta_j\stackrel{\triangle}{=}\frac{\gamma_{th,j}}{\bar{\gamma}(a_j-\gamma_{th,j}(\xi_j+\tilde{\xi}_j+\vartheta_1))}$ notation is made for mathematical tractability. Consequently, by using (\ref{eq:6}), the OP for the $l$th user can be written as
\begin{equation}\label{eq:7}
	\begin{split}
		&P_{out}^l=1-Pr\left(E_{l,1} \cap E_{l,2} \cap \cdots \cap E_{l,l}\right) \\
		&=1-Pr\left(\psi_1>\frac{(\psi_2\bar{\gamma}+\vartheta_2)(\psi_3\bar{\gamma}\vartheta_4+\vartheta_5)\vartheta_3\delta^{\dag}_l}{\bar{\gamma}(\psi_2-\vartheta_2\vartheta_3\delta^{\dag}_l)},\psi_2>\vartheta_2\vartheta_3\delta^{\dag}_l \right), 
	\end{split}
\end{equation}  
where $\delta^{\dag}_l=\underset{1\leq j\leq l}{\max}\left\lbrace \delta_j \right\rbrace$. Note that, (\ref{eq:7}) holds for the condition of $a_j>\gamma_{th,j}(\xi_j+\tilde{\xi}_j+\vartheta_1)$, otherwise the OP results to $1$. Then, (\ref{eq:7}) can be analytically expressed as
\begin{equation}\label{eq:8}
	\begin{split}
		&P_{out}^l=F_{\psi_2}^{(l)}(\vartheta_2\vartheta_3\delta^{\dag}_l)+ \\
		&\int\limits_{y=\vartheta_2\vartheta_3\delta^{\dag}_l}^{\infty}\int\limits_{z=0}^{\infty}F_{\psi_1}\left( \frac{(y\bar{\gamma}+\vartheta_2)(z\bar{\gamma}\vartheta_4+\vartheta_5)\vartheta_3\delta^{\dag}_l}{\bar{\gamma}(y-\vartheta_2\vartheta_3\delta^{\dag}_l)}\right)f_{\psi_3}(z)f_{\psi_2}^{(l)}(y)dzdy, 
	\end{split}
\end{equation}
where $F_{X}^{(l)}$ and $f_{X}^{(l)}$ denote the CDF and PDF of user with the $l$th order statistic. Since channels are considered as i.i.d. Nakagami-$m$ fading, it is well-known that the squared gain of any link will be distributed as Gamma. Thus, corresponding CDFs and PDF of random variables $\psi_1$, $\psi_2$ and $\psi_3$ can be represented as $F_{\psi_1}(x)=1-e^{-xm_{SR}/\hat{\Omega}_{SR}}\sum_{n=0}^{m_{SR}N_S-1}\frac{(xm_{SR}/\hat{\Omega}_{SR})^n}{n!}$, $F_{\psi_2}(x)=1-e^{-xm_{l}/\hat{\Omega}_{l}}\sum_{n_1=0}^{m_{l}N_D-1}\frac{(xm_{l}/\hat{\Omega}_{l})^{n_1}}{n_1!}$ and $f_{\psi_3}(x)=\left(m_{LI}/\Omega_{LI} \right)^{m_{LI}}\frac{x^{m_{LI}-1}}{\Gamma(m_{LI})}e^{-xm_{LI}/\Omega_{LI}}$. Here, $m_{SR}$, $m_{l}$ and $m_{LI}$ denote Nakagami-$m$ channel parameters related to $S-R$, $R-U_l$ and $R-R$ links. If $F_{\psi_1}(x)$ is substituted into (\ref{eq:8}) and then the integrals are rearranged, we obtain
\begin{equation}\label{eq:9}
	\begin{split}
		P_{out}^l&=1-\sum_{n=0}^{m_{SR}N_S-1}\frac{1}{n!}\int_{x=0}^{\infty}e^{-b}f_{\psi_2}^{(l)}(x+\vartheta_2\vartheta_3\delta^{\dag}_l)dx \\
		&\times\underbrace{\int_{z=0}^{\infty}e^{-za}(za+b)^nf_{\psi_3}(z)dz}_{I_1},
	\end{split}
\end{equation}
where $a=\frac{(\bar{\gamma}(x+\vartheta_2\vartheta_3\delta^{\dag}_l)+\vartheta_2)\vartheta_3\vartheta_4\delta^{\dag}_lm_{SR}}{x\hat{\Omega}_{SR}}$ and $b=\frac{a\vartheta_5}{\bar{\gamma}\vartheta_4}$. By using binomial expansion \cite[eq.(1.111)]{Gradshteyn}, integral property given by \cite[eq.(3.381.4)]{Gradshteyn} and $f_{\psi_3}(z)$, $I_1$ is obtained as
\begin{equation}\label{eq:10}
	\begin{split}
		I_1=\sum_{m=0}^{n}\binom{n}{m}\left( \frac{m_{LI}}{\Omega_{LI}}\right) ^{m_{LI}}\frac{\Gamma(m+m_{LI})}{\Gamma(m_{LI})}\frac{b^{n-m}}{a^{-m}}\left(a+\frac{m_{LI}}{\Omega_{LI}} \right)^{-m-m_{LI}}, 
	\end{split}
\end{equation} 
where $\binom{\cdot}{\cdot}$ represents binomial coefficient, and thus (\ref{eq:9}) can be rewritten as
\begin{equation}\label{eq:11}
	\begin{split}
		P_{out}^l&=1-\sum_{n=0}^{m_{SR}N_S-1}\sum_{m=0}^{n}\binom{n}{m}\frac{(m_{LI}/\Omega_{LI})^{m_{LI}}\Gamma(m+m_{LI})}{\Gamma(m_{LI})\Gamma(n+1)} \\
		&\int_{x=0}^{\infty}e^{-b}\frac{b^{n-m}}{a^{-m}}\left(a+\frac{m_{LI}}{\Omega_{LI}} \right)^{-m-m_{LI}}f_{\psi_2}^{(l)}(x+\vartheta_2\vartheta_3\delta^{\dag}_l)dx.
	\end{split}
\end{equation}
In order to proceed, $f_{\psi_2}^{(l)}(x)$ should be determined. By using order statistic properties \cite{DavidHA}, PDF of user with the $l$th order can be expressed as $f_{\psi_2}^{(l)}(x)=Q_l\sum\nolimits_{s=0}^{L-l}\binom{L-l}{s}(-1)^{s}f_{\psi_2}(x)\big(F_{\psi_2}(x) \big)^{l+s-l}$, where $Q_l=L!/((L-l)!(l-1)!)$ \cite{DingZ,JJmen}. If previously defined CDF of $F_{\psi_2}(x)$ and its derivative yielding PDF are substituted into $f_{\psi_2}^{(l)}(x)$, we get
\begin{equation}\label{eq:12}
	\begin{split}
		f_{\psi_2}^{(l)}(x)&=Q_l\sum_{s=0}^{L-l}\sum_{s_1=0}^{l+s-1}\sum_{n_1=0}^{s_1(m_lN_D-1)}\binom{L-l}{s}\binom{l+s-1}{s_1}(-1)^{s+s_1} \\
		&\frac{(m_l/\hat{\Omega}_l)^{m_lN_D}}{\Gamma(m_lN_D)}\theta_{n_1}(s_1,m_lN_D)x^{n_1+m_lN_D-1}e^{-\frac{xm_l(s_1+1)}{\hat{\Omega}_l}}. 
	\end{split}
\end{equation}          
In order to obtain the closed-form of $f_{\psi_2}^{(l)}(x)$ in (\ref{eq:12}), binomial expansion \cite[eq.(1.111)]{Gradshteyn} and power series method given by \cite[eq.(0.314)]{Gradshteyn} are applied to $\big(F_{\psi_2}(x) \big)^{l+s-l}$. Here, $\theta_{n_1}(s_1,m_lN_D)$ represents multinomial coefficient consisting of a recursive summation \cite{Mtoka}. Finally, by substituting (\ref{eq:12}) into (\ref{eq:11}) and applying algebraic manipulations, the OP corresponding to the $l$th user can be obtained as 
\begin{equation}\label{eq:13}
	\begin{split}
		P_{out}^l&= 1-Q_l\sum_{n=0}^{m_{SR}N_S-1}\sum_{m=0}^{n}\sum_{s=0}^{L-l}\sum_{s_1=0}^{l+s-1}\sum_{n_1=0}^{s_1(m_lN_D-1)}\sum_{n_2=0}^{n_1+m_lN_D-1}\sum_{n_3=0}^{n}\binom{n}{m}\binom{L-l}{s}\binom{l+s-1}{s_1}\binom{n_1+m_lN_D-1}{n_2} \\
		&\binom{n}{n_3}(-1)^{s+s_1}\frac{\Gamma(m+m_{LI})(m_{LI}/\Omega_{LI})^{m_{LI}}(m_l/\hat{\Omega}_l)^{m_lN_D}}{\Gamma(n+1)\Gamma(m_{LI})\Gamma(m_{l}N_D)}\theta_{n_1}(s_1,m_lN_D)(\vartheta_2\vartheta_3\delta^{\dag}_l)^{n_1+m_lN_D-n_2-1} \\
		&e^{-\frac{(\vartheta_2\vartheta_3\delta^{\dag}_l)m_l(s_1+1)}{\hat{\Omega}_l}-\frac{\vartheta_3\vartheta_5\delta^{\dag}_lm_{SR}}{\hat{\Omega}_{SR}}}\left(\frac{\bar{\gamma}\vartheta_4}{\vartheta_5} \right)^m\left(\frac{\vartheta_3\vartheta_5\delta^{\dag}_lm_{SR}}{\hat{\Omega}_{SR}} \right)^n\left(\vartheta_2\vartheta_3\delta^{\dag}_l+\frac{\vartheta_2}{\bar{\gamma}} \right)^{n-n_3}\left( \Omega_{LI}\hat{\Omega}_{SR}\right) ^{m+m_{LI}} \\
		&\left(\bar{\gamma}\vartheta_3\vartheta_4\delta^{\dag}_lm_{SR}\Omega_{LI}+m_{LI}\hat{\Omega}_{SR} \right)^{-m-m_{LI}}\Theta_l(x).
	\end{split}
\end{equation}

In (\ref{eq:13}), $\Theta_l(x)$ can be expressed by
\setcounter{equation}{13}
\begin{equation}\label{eq:14}
	\begin{split}
		\Theta_l(x)&=\int\limits_{x=0}^{\infty}x^{n_2+n_3+m+m_{LI}-n}e^{-\frac{xm_{l}\left( s_1+1\right) }{\hat{\Omega}_l}-\frac{\left( \vartheta_2\vartheta_3\delta^{\dag}_l+\vartheta_2/\bar{\gamma}\right) \vartheta_3\vartheta_5\delta^{\dag}_lm_{SR}}{x\hat{\Omega}_{SR}}} \\
		&\left(x+\frac{\vartheta_4(\bar{\gamma}\vartheta_2\vartheta_3\delta^{\dag}_l+\vartheta_2)\vartheta_3\delta^{\dag}_lm_{SR}\Omega_{LI}}{\bar{\gamma}\vartheta_3\vartheta_4\delta^{\dag}_lm_{SR}\Omega_{LI}+m_{LI}\hat{\Omega}_{SR}} \right)^{-m-m_{LI}}dx. 
	\end{split}
\end{equation}
Unfortunately, the closed-form of the integral in (\ref{eq:14}) does not exist.

\subsection{Lower Bound Analysis}

In order to find a tight approximation for the exact OP given by (\ref{eq:13}), SIDNR given in (\ref{eq:4}) can be upper-bounded, thus a tight lower-bound for the exact OP can be obtained. Firstly, (\ref{eq:7}) can be rewritten approximately as
\begin{equation}\label{eq:15}
	P_{out}^l\approx1-Pr\left(\frac{W\frac{1}{\vartheta_4}\bar{\gamma}\psi_2\frac{1}{\vartheta_2}}{W\frac{1}{\vartheta_4}+\bar{\gamma}\psi_2\frac{1}{\vartheta_2}}>\bar{\gamma}\vartheta_3\delta^{\dag}_l \right), 
\end{equation} 
where $W=\bar{\gamma}\psi_1/(\bar{\gamma}\psi_3+\vartheta_5/\vartheta_4)$ for mathematical simplicity. Then, by using the harmonic mean property of two random variables defined as $xy/(x+y)\leq\min(x,y)$, lower-bound for the exact OP can be analytically expressed as 
\begin{equation}\label{eq:16}
	\begin{split}
		P_{out}^{l,low}&=1-
		Pr\left(\min\left(W\frac{1}{\vartheta_4},\bar{\gamma}\psi_2\frac{1}{\vartheta_2} \right)  >\bar{\gamma}\vartheta_3\delta_l^{\dag}\right)  \\
		&=1-\overline{F}_{W}\left(\bar{\gamma}\vartheta_3\vartheta_4\delta_l^{\dag} \right)\overline{F}_{\psi_2}^{(l)}\left(\vartheta_2\vartheta_3\delta_l^{\dag} \right).  
	\end{split}
\end{equation} 

In (\ref{eq:16}), the CDF of $F_W(x)$ can be mathematically expressed as
\begin{equation}\label{eq:17}
	\begin{split}
		F_W(x)&=Pr\bigg(\frac{\bar{\gamma}\psi_1}{\bar{\gamma}\psi_3+(\vartheta_5/\vartheta_4)}\leq\underbrace{\bar{\gamma}\vartheta_3\vartheta_4\delta_l^{\dag}}_{x} \bigg) \\
		&=Pr\left(\psi_1\leq x\left(\psi_3+\frac{\vartheta_5}{\bar{\gamma}\vartheta_4} \right)  \right) \\
		&=1-\int\limits_{y=0}^{\infty}\int\limits_{x=x\left(y+\frac{\vartheta_5}{\bar{\gamma}\vartheta_4} \right) }^{\infty}f_{\psi_1}(x)f_{\psi_3}(y)dxdy.  
	\end{split}
\end{equation}
Then, by substituting previously defined PDF of $f_{\psi_3}(x)$ and derivative of the CDF $F_{\psi_1}(x)$ into (\ref{eq:17}), we can obtain
\begin{equation}\label{eq:18}
	\begin{split}
		&F_W(x)=1-\sum_{n=0}^{m_{SR}N_S-1}\frac{1}{n!}\left(\frac{m_{LI}}{\Omega_{LI}} \right)^{m_{LI}}\frac{1}{\Gamma(m_{LI})} \\
		&\int\limits_{y=0}^{\infty}y^{m_{LI}-1}\left(x\left( y+\frac{\vartheta_5}{\bar{\gamma}\vartheta_4}\right) \frac{m_{SR}}{\hat{\Omega}_{SR}} \right)^{n}e^{-x\left( y+\frac{\vartheta_5}{\bar{\gamma}\vartheta_4}\right) \frac{m_{SR}}{\hat{\Omega}_{SR}}-y\frac{m_{LI}}{\Omega_{LI}}}dy. 
	\end{split}
\end{equation}
Finally, with the help of integral property given by \cite[eq.(3.381.4)]{Gradshteyn}, the CDF of $F_W(x)$ can be derived as    
\begin{equation}\label{eq:19}
	\begin{split}
		&F_W(x)=1-\sum_{n=0}^{m_{SR}N_S-1}\sum_{n_2=0}^{n}\binom{n}{n_2}\frac{(m_{LI}/\Omega_{LI})^{m_{LI}}(m_{SR}/\hat{\Omega}_{SR})^n}{\Gamma(n+1)\Gamma(m_{LI})} \\
		&\Gamma(n_2+m_{LI})\left(\frac{\vartheta_5}{\bar{\gamma}\vartheta_4} \right)^{n-n_2}x^n\left(\frac{xm_{SR}}{\hat{\Omega}_{SR}}+\frac{m_{LI}}{\Omega_{LI}} \right)^{-n_2-m_{LI}}e^{-\frac{x\vartheta_5m_{SR}}{\bar{\gamma}\vartheta_4\hat{\Omega}_{SR}}}. 
	\end{split}
\end{equation}
On the other hand, if the PDF of $f_{\psi_2}^{(l)}(x)$ given in (\ref{eq:12}) is integrated with respect to $x$, the CDF of $F_{\psi_2}^{(l)}(x)$ is obtained as
\begin{equation}\label{eq:20}
	\begin{split}
		F_{\psi_2}^{(l)}(x)&=1-Q_l\sum_{s=0}^{L-l}\sum_{s_1=1}^{l+s}\sum_{n_1=0}^{s_1(m_lN_D-1)}\binom{L-l}{s}\binom{l+s}{s_1} \\
		&\frac{(-1)^{s+s_1-1}}{l+s}\theta_{n_1}(s_1,m_lN_D)x^{n_1}e^{-\frac{xm_ls_1}{\hat{\Omega}_l}}. 
	\end{split}
\end{equation}
Afterwards, complementary versions of CDFs given in (\ref{eq:19}) and (\ref{eq:20}) are substituted into (\ref{eq:16}), a tight lower-bound of the exact OP corresponding to $l$th user can be obtained in closed-form.

\subsection{Asymptotic Analyses}
In order to reveal further insights for the system performance, asymptotic behavior of OP is considered by applying high SNR approximation in this subsection. Therefore, we have carried out the analyses according to two cases which are presented in the following subsections.

\subsubsection{Under Ideal Conditions}

\paragraph*{$\bullet$ When the quality of LI cancellation is $\mu\neq 1$}
In the presence of ideal conditions (which also means that there are no CEEs in the first and second hops), system exploits benefits of diversity order and array gain at high SNR values (when $\bar{\gamma}\rightarrow\infty$). Thus, (\ref{eq:15}) can be approximated as
\begin{equation}\label{eq:21}
\begin{split}
P_{out}^{l,\infty}&\approx1-Pr\left(\frac{W\vartheta_1^{'}\bar{\gamma}\psi_2\vartheta_2^{'}}{W\vartheta_1^{'}+\bar{\gamma}\psi_2\vartheta_2^{'}}>\bar{\gamma}\vartheta_1^{'}\vartheta_2^{'}\delta^{\dag}_l \right) \\
&=1-
Pr\left(\min\left(W\vartheta_1^{'},\bar{\gamma}\psi_2\vartheta_2^{'} \right)  >\bar{\gamma}\vartheta_1^{'}\vartheta_2^{'}\delta^{\dag}_l\right) \\
&=F_{W}^{\infty}\left( \bar{\gamma}\vartheta_2^{'}\delta^{\dag}_l\right) +F_{\psi_2}^{(l),\infty}\left(\vartheta_1^{'}\delta^{\dag}_l\right), 
\end{split}
\end{equation}
where $W\simeq \psi_1/\psi_3$, $\vartheta_1^{'}=1+\kappa_{SR}^2$ and $\vartheta_2^{'}=1+\kappa_{RU}^2$ are defined. By using high SNR approximation approach \cite{ZWang}, we can express asymptotic OP in (\ref{eq:21}) as in the form $P_{out}^{l,\infty}\approx\left(AG\bar{\gamma} \right)^{-DO}+\textit{O}\left( \bar{\gamma}^{-DO}\right)$, where $AG$ is the array gain, $DO$ is the diversity order and $\textit{O}(\cdot)$ represents high order terms to be neglected. Firstly, the asymptotic CDF of $W$ can be derived by $F_{W}^{\infty}(x)=Pr(\psi_1/\psi_3\leq x)=\int_{y=0}^{\infty}F_{\psi_1}^{\infty}(yx)f_{\psi_3}(y)dy$. Here, the CDF of $\psi_1$ is expressed as $F_{\psi_1}(x)=\frac{\gamma(m_{SR}N_S,xm_{SR}/\Omega_{SR})}{\Gamma(m_{SR}N_S)}$ in terms of lower incomplete Gamma function \cite[eq.(8.350.1)]{Gradshteyn}, then it can be asymptotically obtained as $F_{\psi_1}^{\infty}(x)\approx\frac{(xm_{SR}/\Omega_{SR})^{m_{SR}N_S}}{\Gamma(m_{SR}N_S+1)}$ by using the property of $\gamma(x,y\rightarrow 0)\approx y^x/x$ \cite[eq.(45:9:1)]{Oldham}. If $F_{\psi_1}^{\infty}(x)$ and previously defined $f_{\psi_3}(y)$ are substituted into $F_{W}^{\infty}(x)$ by replacing $x$ with $\bar{\gamma}\vartheta_2^{'}\delta^{\dag}_l$, we obtain $F_{W}^{\infty}(\bar{\gamma}\vartheta_2^{'}\delta^{\dag}_l)=(\chi_1\bar{\gamma})^{-(1-\mu)m_{SR}N_S}$, where $\chi_1$ can be obtained as
\begin{equation}\label{eq:22}
	\chi_1=\left(\frac{\Gamma(m_{SR}N_S+m_{LI})}{\Gamma(m_{SR}N_S+1)\Gamma(m_{LI})}\left(\frac{\vartheta_2^{'}\varLambda_l^{\dag}m_{SR}}{\Omega_{SR}m_{LI}} \right)^{m_{SR}N_S}  \right)^{-\frac{1}{(1-\mu)m_{SR}N_S}}.
\end{equation}
In (\ref{eq:22}), $\varLambda_l^{\dag}=\bar{\gamma}\delta^{\dag}_l$ and independent from $\bar{\gamma}$. Therefore, the exponentially dominant constant on the average SNR ($\bar{\gamma}$) within the expression of $F_{W}^{\infty}(\bar{\gamma}\vartheta_2^{'}\delta^{\dag}_l)$ equals to $(1-\mu)m_{SR}N_S$ for the first hop. Asymptotic expression of $F_{\psi_2}^{(l),\infty}(x)$ can be derived by taking into account of lower order terms related to variable of $x$ in (\ref{eq:20}) as $F_{\psi_2}^{(l),\infty}(x)\approx\binom{L}{l}\left(\frac{(xm_l/\Omega_l)^{m_lN_D}}{\Gamma(m_lN_D+1)} \right)^l$. Afterwards, by replacing $x$ with $\vartheta_1^{'}\delta^{\dag}_l$ and after mathematical manipulations, we obtain $F_{\psi_2}^{(l),\infty}(\vartheta_1^{'}\delta^{\dag}_l)=(\chi_2\bar{\gamma})^{-m_lN_Dl}$, where $\chi_2$ can be found as
\begin{equation}\label{eq:23}
	\chi_2=\left(\binom{L}{l}\frac{1}{(\Gamma(m_lN_D+1))^l} \right)^{-\frac{1}{m_lN_Dl}}\frac{\Omega_l}{\vartheta_1^{'}\varLambda_l^{\dag}m_l}.
\end{equation}
From (\ref{eq:23}), the exponentially dominant constant on the average SNR ($\bar{\gamma}$) within the expression of $F_{\psi_2}^{(l),\infty}(\vartheta_1^{'}\delta^{\dag}_l)$ equals to $m_lN_Dl$ for the second hop. Consequently, if $F_{W}^{\infty}\left( \bar{\gamma}\vartheta_2^{'}\delta^{\dag}_l\right)$ and $F_{\psi_2}^{(l),\infty}\left(\vartheta_1^{'}\delta^{\dag}_l\right)$ are substituted into (\ref{eq:21}), and with the help of asymptotic form $P_{out}^{l,\infty}\approx\left(AG\bar{\gamma} \right)^{-DO}+\textit{O}\left( \bar{\gamma}^{-DO}\right)$, the asymptotic OP of the $l$th user can be obtained in simple form with diversity order metric $DO=\min\left\lbrace \left(1-\mu\right)m_{SR}N_S,m_{l}N_Dl\right\rbrace$. On the other hand, array gain can be found by using (\ref{eq:22}) and (\ref{eq:23}) as  
\begin{equation}\label{eq:24}
	AG=\left\{
	\begin{array}{ll}
		\chi_1 & (1-\mu)m_{SR}N_S<m_{l}N_Dl \\
		\chi_2 & (1-\mu)m_{SR}N_S>m_{l}N_Dl \\
		\chi_1+\chi_2 & (1-\mu)m_{SR}N_S=m_{l}N_Dl \\
	\end{array}
	\right..
\end{equation}                   
\paragraph*{$\bullet$ When the quality of LI cancellation is $\mu=1$}
In this case, $S-R$ link will be extremely dominant in $e2e$ SIDNR due to the high LI effect. Therefore, asymptotic OP for the $l$th user can be expressed as $P_{out}^{l,\infty}\approx F_W(\vartheta_2^{'}\varLambda_l^{\dag})$, where $W\approx\psi_1/\psi_3$, by neglecting the effect of $R-U_l$ link for high SNR values. Since $F_W(\vartheta_2^{'}\varLambda_l^{\dag})$ is independent from average SNR which also yields error floor level at high SNR values (also means zero diversity), we can not carry out high SNR approximation provided in \cite{ZWang}. Therefore, $F_W(x)$ can be obtained as
\begin{equation}\label{eq:25}
	\begin{split}
		F_W(x)&=Pr\bigg(\frac{\psi_1}{\psi_3}\leq\underbrace{\vartheta_2^{'}\varLambda_l^{\dag}}_{x} \bigg) \\
		&=1-\int_{y=0}^{\infty}\int_{x=xy}^{\infty}f_{\psi_1}(x)f_{\psi_2}(y)dxdy. 
	\end{split}
\end{equation}
Then, by substituting the CDF of $\psi_1$ to get rid of the inner integral and PDF of $\psi_3$ into (\ref{eq:25}), and with the help of integral property provided by \cite[eq.(3.381.4)]{Gradshteyn}, the asymptotic OP of the $l$th user can be derived as
\begin{equation}\label{eq:26}
	\begin{split}
		P_{out}^{l,\infty}&=1-\sum_{n=0}^{m_{SR}N_S-1}\frac{(m_{LI}/\Omega_{LI})^{m_{LI}}(m_{SR}/\Omega_{SR})^n}{\Gamma(n+1)\Gamma(m_{LI})} \\
		&\Gamma(n+m_{LI})(\vartheta_2^{'}\varLambda_l^{\dag})^n\left(\frac{\vartheta_2^{'}\varLambda_l^{\dag}m_{SR}}{\Omega_{SR}}+\frac{m_{LI}}{\Omega_{LI}} \right)^{-n-m_{LI}} 
	\end{split}
\end{equation}

\subsubsection{Under Practical Conditions}
Since CEE parameters $\sigma_{e,l}^2$ and $\sigma_{e,SR}^2$ are dominant on the $e2e$ SIDNR, we can not apply asymptotic property $\gamma(x,y\rightarrow 0)\approx y^x/x$ given in \cite[eq.(45:9:1)]{Oldham}. Thus, by considering the dominance of CEE effects, in case of all quality of LI cancellation values ($\mu$), the predefined constants of $\vartheta_2$ and $\vartheta_5$ given by (\ref{eq:5}) can be approximated as $\vartheta_2\approx\bar{\gamma}\sigma_{e,l}^2$ and $\vartheta_5\approx\bar{\gamma}\sigma_{e,SR}^2$, respectively. By substituting $\vartheta_2$ and $\vartheta_5$ together with other constants given by (\ref{eq:5}) into (\ref{eq:13}), asymptotic OP of $l$th user in the presence of CEEs can be obtained.

\section{Numerical Results}
In this section, theoretical results for the investigated system verified by Monte Carlo simulations are presented. An exemplary, the scenario with three mobile users ($L=3$) is considered. Unless otherwise stated, markers illustrate simulation results, $SNR=\bar{\gamma}=P/\sigma^2$, $\alpha=3$ (for urban area cellular radio), $\lambda=1$ as in \cite{Duarte,Rodriguez}. While power coefficients to be allocated to users are set as $a_1=1/2$, $a_2=1/3$ and $a_3=1/6$, target SIDNR thresholds related to mobile users for FD transmission are determined as $\gamma_{th,1}=0.9$, $\gamma_{th,2}=1.5$ and $\gamma_{th,3}=2$, respectively. Also, normalized distances of $S-R$ link and $R-U_l$ links are fixed as $d_{SR}=0.5$ and $d_{1}=d_{2}=d_{3}=0.5$, respectively. For simplicity, Nakagami-$m$ channel parameters of $R-U_l$ links are assumed as $m_{1}=m_{2}=m_{3}=m_{RU}$. For easy of reading, lower bound (LB), asymptotic (Asymp) and theoretical (Theo) abbreviations are made. 

\begin{figure}[!b]
	\centering
	\includegraphics[width=0.70\columnwidth]{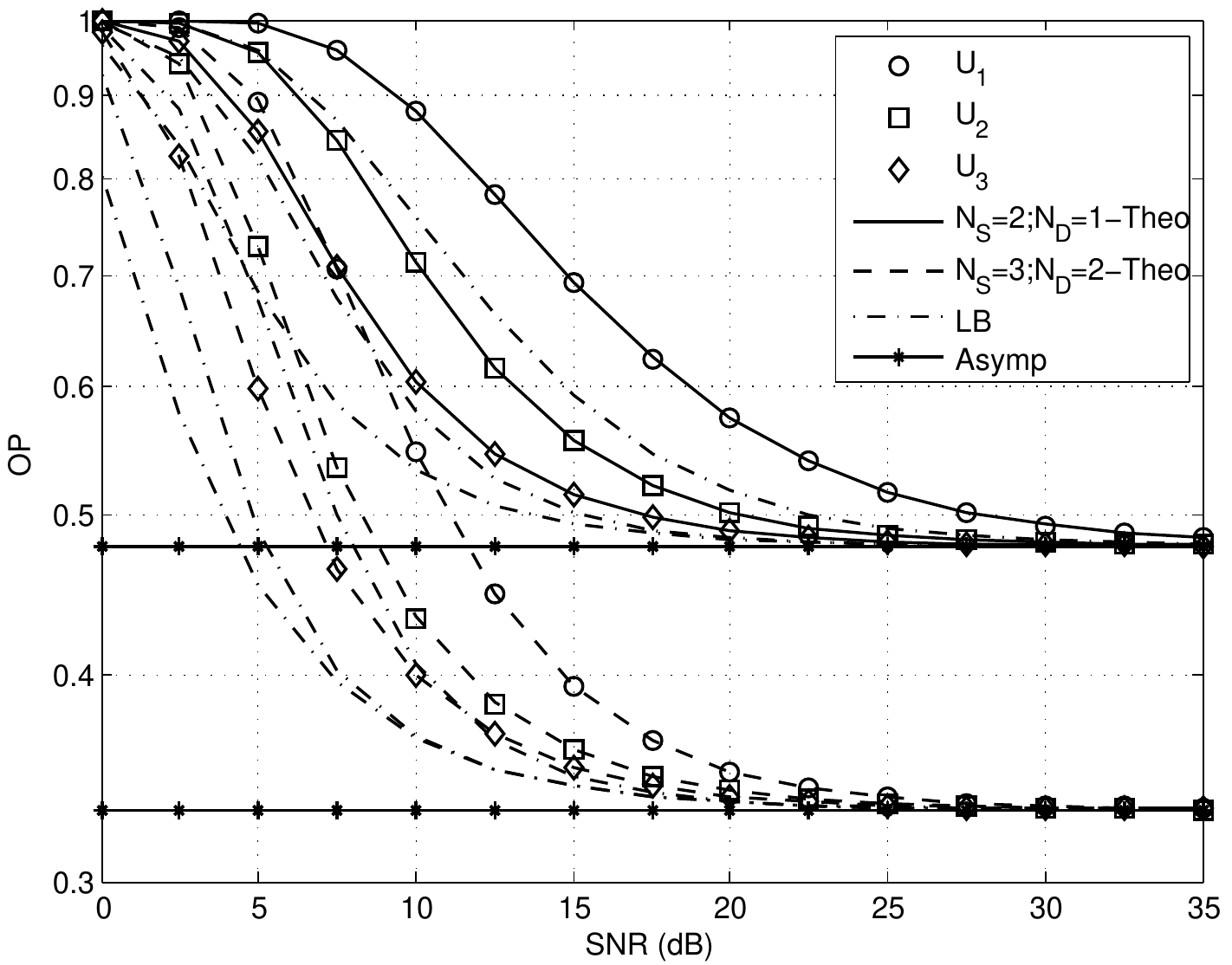}
	\caption{OP of the investigated system in case of $\mu=1$ and ideal conditions.}
	\label{fig:1}
	\vspace{-2 mm}
\end{figure}  
Fig. 1 depicts OP curves of the investigated system in case of $\mu=1$ (which means the worst scenario of LI cancellation process), $m_{SR}=m_{LI}=m_{RU}=1$, $\sigma_{ipsic}^2=0$, $\kappa_{SR}=\kappa_{RU}=0$ under ideal channel conditions ($\sigma_{e,SR}^2=0$ and $\sigma_{e,l}^2=0$) for different antenna configurations. We observe from the figure that OP performance of the investigated system is strictly limited by error floor level, which is also known as zero-diversity and validated by asymptotic results, at high SNR values for all users. Also, OPs of all users are exposed to the same level regardless of the number of antennas. On the other hand, given the increased number of antennas (when the configurations $N_S=3$; $N_D=2$ and $N_S=2$; $N_D=1$ are compared), performance of the system can be improved in the low SNR region and with the decrease of error floor level in the high SNR region. In addition, we also observed that the exact results are supported by LB curves which are quite tight and match well in the high SNR region. 

Fig. 2 illustrates OP performance of the system for different antenna configurations for $\mu=0.2$, $m_{SR}=m_{LI}=m_{RU}=1$, $\sigma_{e,SR}^2=\sigma_{e,l}^2=0$, $\sigma_{ipsic}^2=0$ and $\kappa_{SR}=\kappa_{RU}=0$. As clearly observed from the figure, there is no error floor level when $\mu\neq 1$, and thus all users enjoy benefits of diversity order in the high SNR region. This result is also verified by the asymptotic curves which are obtained by theoretical analyses. Particularly, according to results of the first user, diversity orders are $(1-\mu)m_{SR}N_S$ for $N_S=1$; $N_D=1$ and $N_S=2$; $N_D=2$ configurations, and $m_{RU}N_Dl$ for $N_S=3$; $N_D=2$ configuration, respectively. Also, if $N_S=3$; $N_D=2$ and $N_S=2$; $N_D=2$ configurations are compared for an OP value of $10^{-5}$, $11$ dB more SNR gain can be achieved for the second and third users while $7$ dB for the first user, which implies that MRT beamforming is much more effective on the performance of the second and third users than first user. 

\begin{figure}[!b]
	\centering
	\includegraphics[width=0.70\columnwidth]{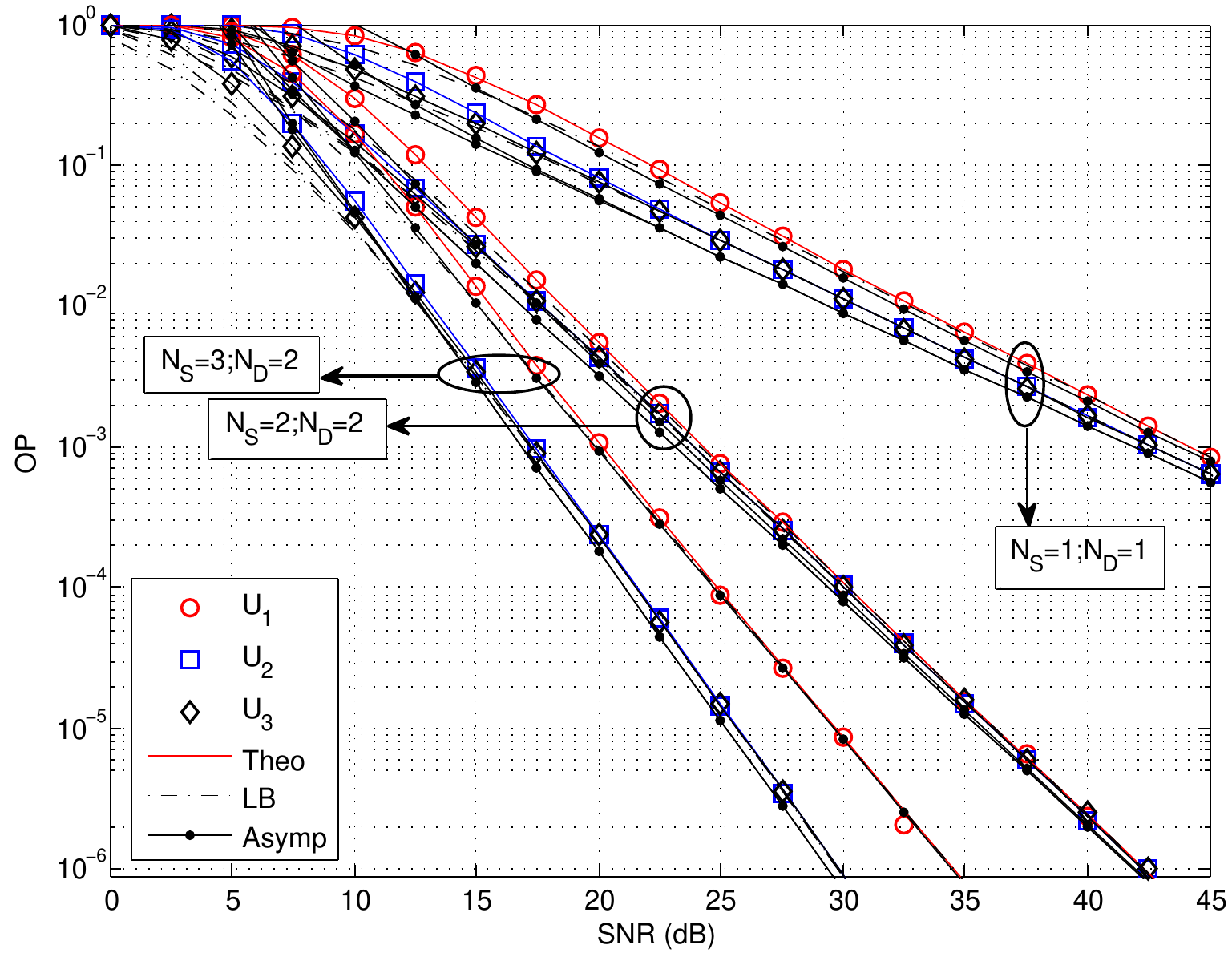}
	\caption{OP of the investigated system in case of $\mu=0.2$ and different number of transmit/receive antenna configurations under ideal conditions.}
	\label{fig:2}
	\vspace{-2 mm}
\end{figure} 

In Fig. 3, OP performance of the system is presented for different RHIs parameters and antenna configurations. All curves are obtained for $\mu=0.2$, $\sigma_{e,SR}^2=\sigma_{e,l}^2=0$, $\sigma_{ipsic}^2=0$ and $m_{SR}=m_{LI}=m_{RU}=1$. From the figure, we observed that RHIs highly deteriorate the performance of users if $\kappa_{SR}=\kappa_{RU}=0$ and $\kappa_{SR}=\kappa_{RU}=0.16$ configurations are compared. Particularly, according to $N_S=3$; $N_D=2$ results, performance gap between $\kappa_{SR}=\kappa_{RU}=0$ and $\kappa_{SR}=\kappa_{RU}=0.16$ in terms of the second and third users is approximately $15$dB for an OP value of $10^{-5}$ while $12.5$ dB in terms of the first user. Similar results are obtained for $N_S=1$; $N_D=1$ configuration. This result reveals that the impact of RHIs is more effective on the performance of the second and third users relative to the first user. On the other hand a significant performance gain can be achieved as the number of antennas is increased, even under the effect of RHIs. Moreover, diversity order of the system is not effected by RHIs and asymptotic analysis also validates this observation. 
\begin{figure}[!b]
	\centering
	\includegraphics[width=0.70\columnwidth]{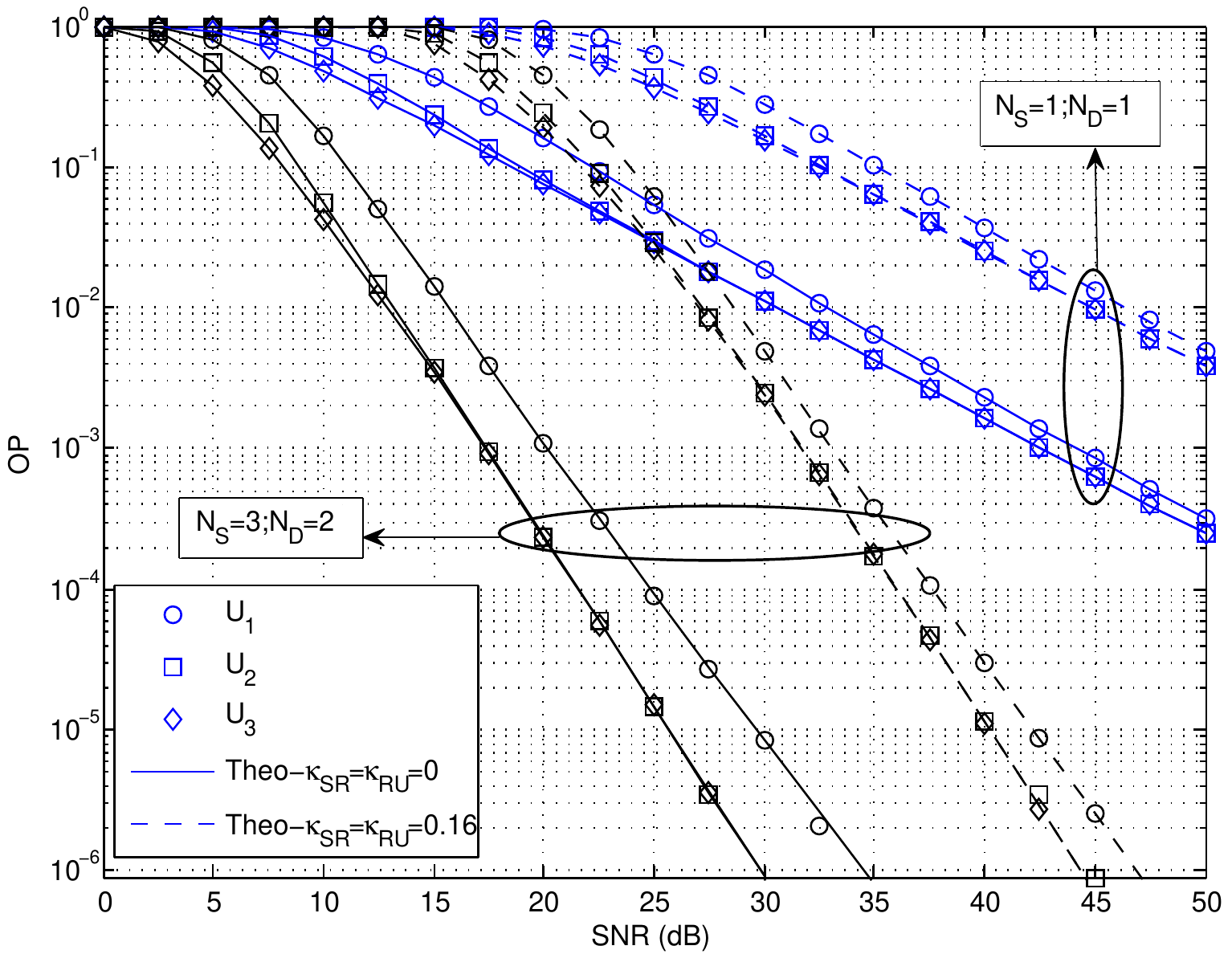}
	\caption{OP of the investigated system in case of $\mu=0.2$ and different number of transmit/receive antenna configurations and RHIs.}
	\label{fig:3}
	\vspace{-2 mm}
\end{figure}   

Figs. 4 and 5 depict OP curves of three users in the presence of practical channel conditions ($\sigma_{e,SR}^2=\sigma_{e,l}^2=0.03$) for $\mu=0.2$, $\sigma_{ipsic}^2=0$ and $\kappa_{SR}=\kappa_{RU}=0$. In both figures, antenna configurations are set as $N_1$:$(N_S=1;N_D=2)$, $N_2$:$(N_S=2;N_D=1)$ and $N_3$:$(N_S=2;N_D=2)$ while channel parameters are $m_1$:$(m_{SR}=1;m_{LI}=1;m_{RU}=2)$ and $m_2$:$(m_{SR}=2;m_{LI}=1;m_{RU}=1)$. In both figures, we observe error floor levels in the high SNR region caused by the effects of CEEs, even $\mu\neq 1$. From Fig. 4, if we compare $N_1$ (also means MRC) and $N_2$ (also means MRT) configurations, MRT is better than MRC when the channel condition in the second hop is better ($m_1$ conf.), while MRC is better than MRT when the channel condition in the first hop is better ($m_2$ conf.). Consequently, performance behavior of the first user according to MRT and MRC schemes also depend on the channel conditions. Also, similar observations are obtained for the second and third users from Fig. 5. Furthermore, as clearly seen in both figures, although MRT and MRC schemes improve the performance of the system, hybrid scheme of MRT/MRC significantly increases OP performance of all users. Note also that hybrid scheme performs better in the low SNR region, significantly better in the high SNR region for $m_1$ configuration.

\begin{figure}[!t]
	\centering
	\includegraphics[width=0.70\columnwidth]{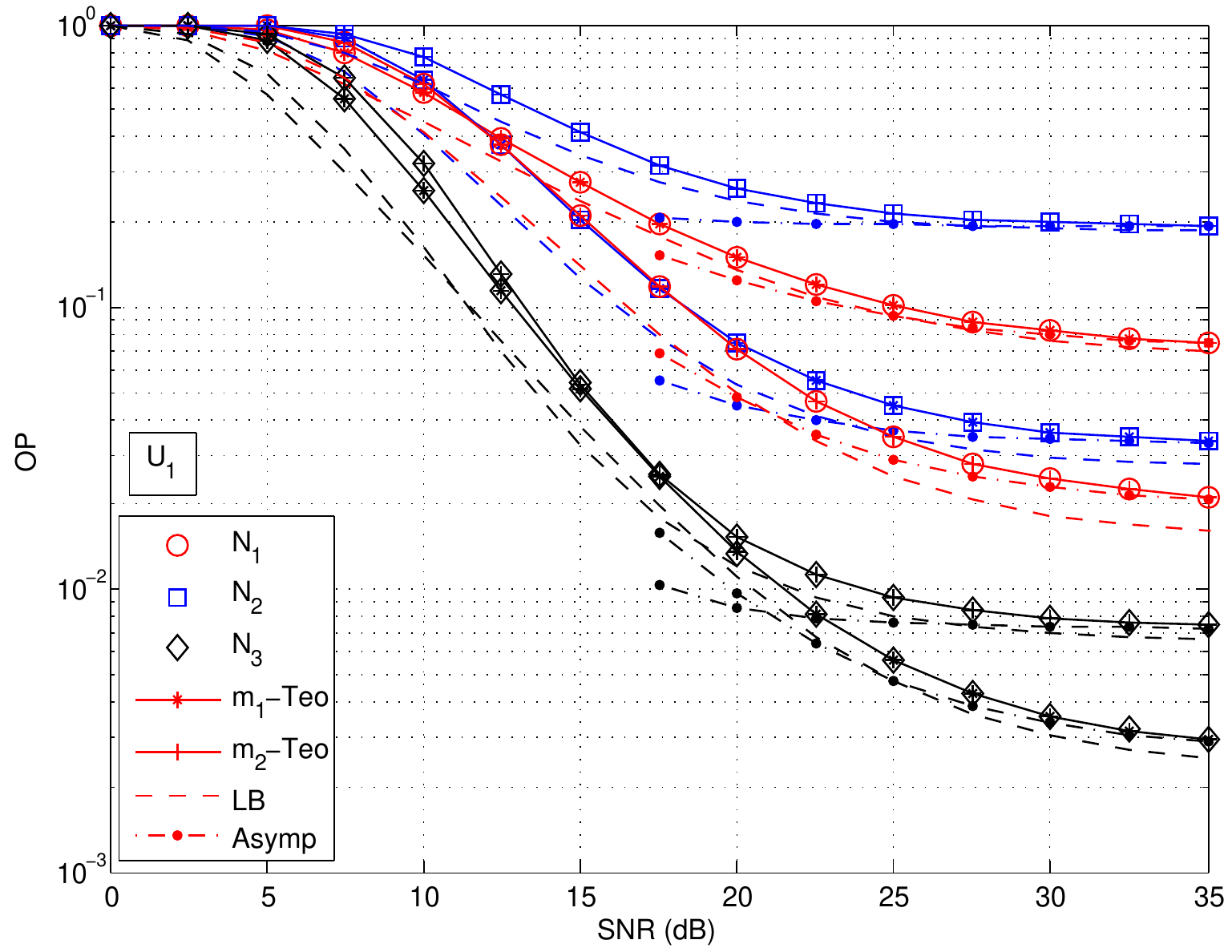}
	\caption{OP of the first user $U_1$ in case of $\mu=0.2$, different fading and number of antenna configurations in the presence of CEEs and ipSIC.}
	\label{fig:4}
	\vspace{-2 mm}
\end{figure}
\begin{figure}[!t]
	\centering
	\includegraphics[width=0.70\columnwidth]{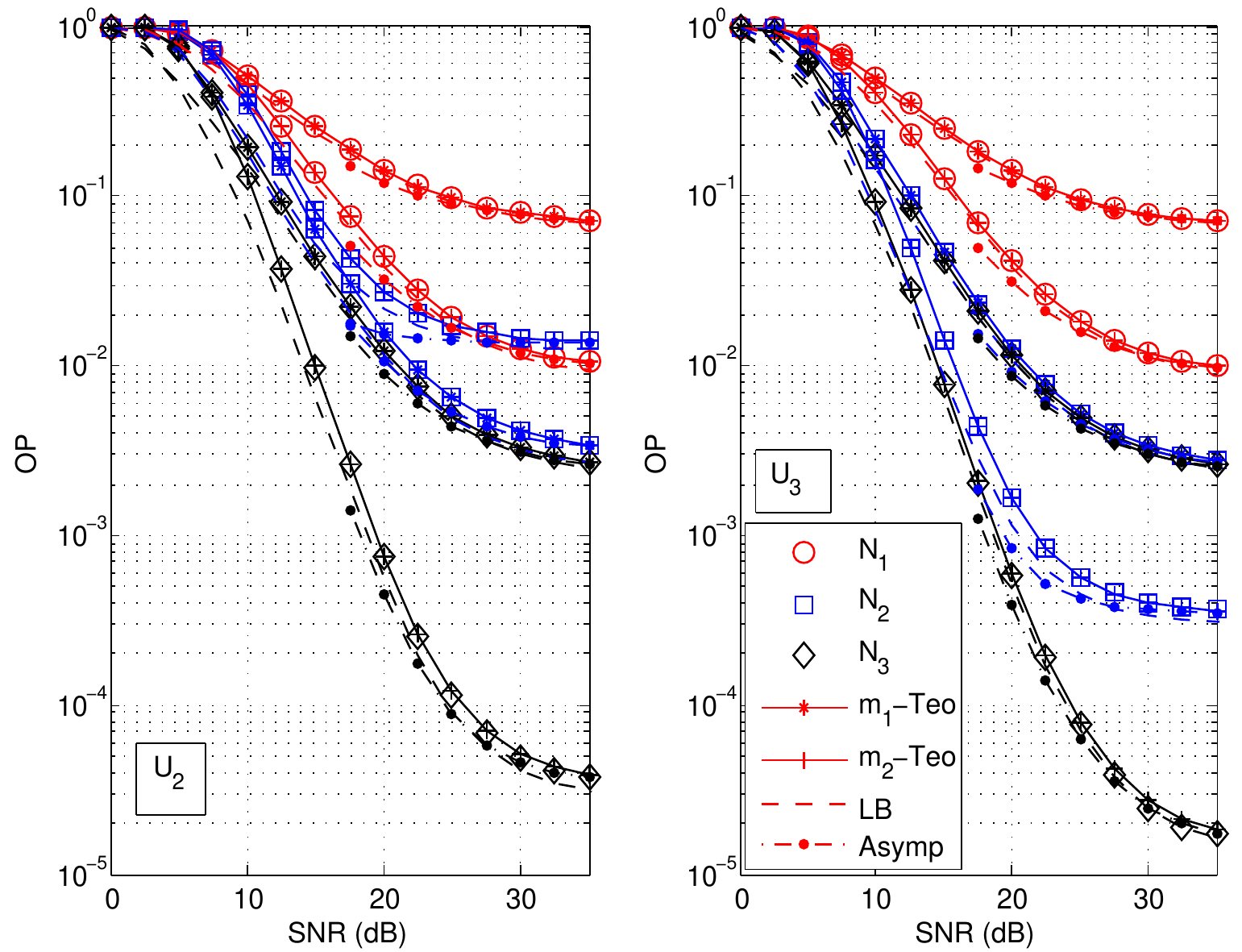}
	\caption{OP of the second $U_2$ and third $U_3$ users in case of $\mu=0.2$, different fading and number of antenna configurations in the presence of CEEs and ipSIC.}
	\label{fig:5}
	\vspace{-2 mm}
\end{figure}
\begin{figure}[!t]
	\centering
	\includegraphics[width=0.68\columnwidth]{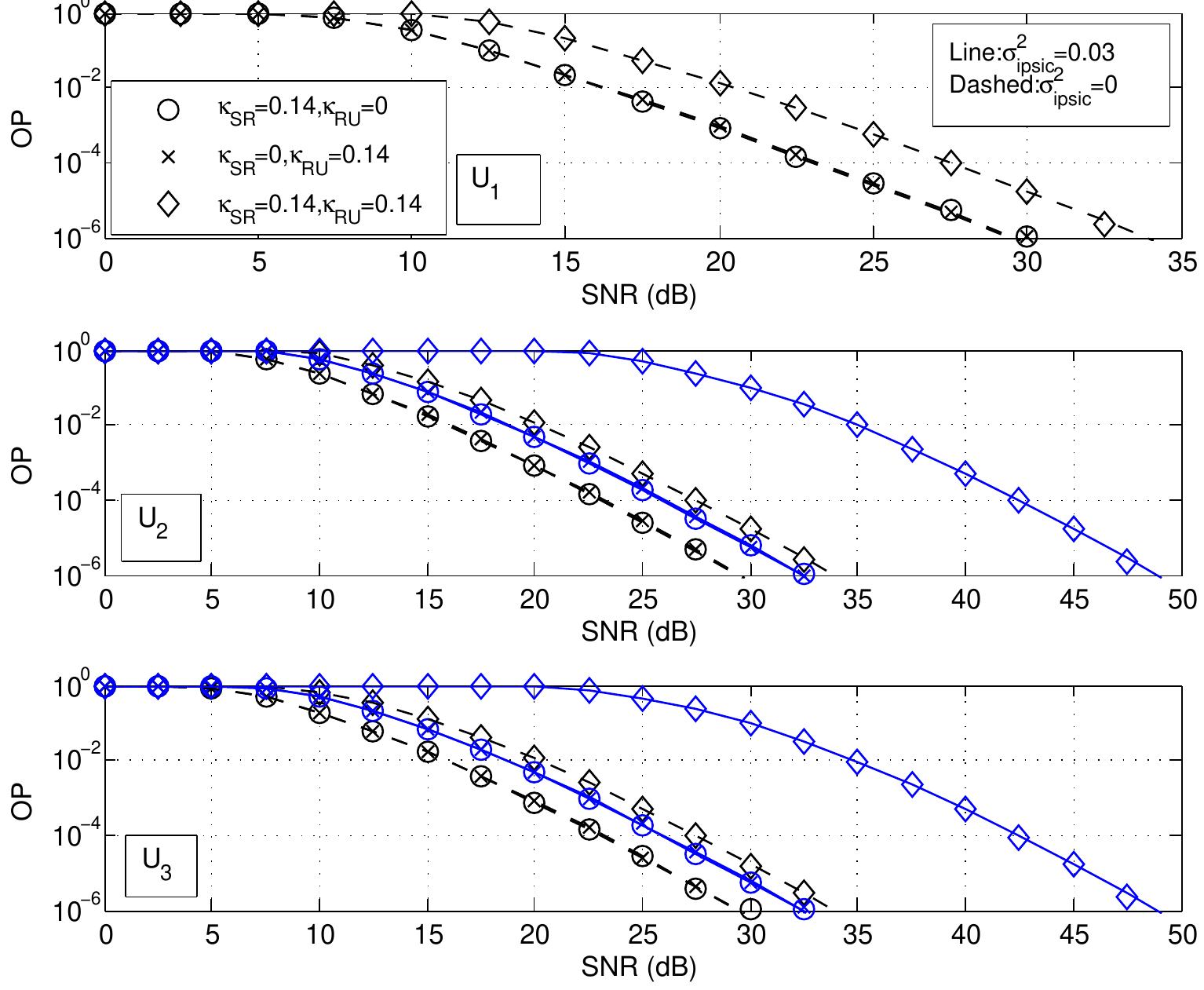}
	\caption{OP of users in case of $\mu=0.2$ and different parameters of RHIs and ipSIC.}
	\label{fig:6}
	\vspace{-3 mm}
\end{figure}

In Fig. 6, OP curves of all users are individually depicted for different RHIs and ipSIC parameters in case of $\mu=0.2$, $\sigma_{e,SR}^2=\sigma_{e,l}^2=0$, ($N_2=2$; $N_D=2$), ($m_{SR}=2$; $m_{LI}=1$; $m_{RU}=2$) configurations. It is obvious that the first user is not effected by ipSIC since it does not perform SIC cancellation. However, RHIs seriously deteriorate the performance of all users. In addition, ($\kappa_{SR}=0.14$; $\kappa_{RU}=0$) and ($\kappa_{SR}=0$; $\kappa_{RU}=0.14$) configurations exhibit the same OP performance, thus RHIs in the first and second hops have the same effect on the system performance. From results of the second user, for an OP value of $10^{-4}$, difference between $\sigma_{ipsic}^2=0.03$ and $\sigma_{ipsic}^2=0$ is approximately $15$ dB in case of ($\kappa_{SR}=0.14$; $\kappa_{RU}=0.14$), while $3$ dB in case of ($\kappa_{SR}=0.14$; $\kappa_{RU}=0$). Similar results can be obtained for the third user. This observation reveals that RHIs effect the performance more than ipSIC.   
\begin{figure}[!b]
	\centering
	\includegraphics[width=0.70\columnwidth]{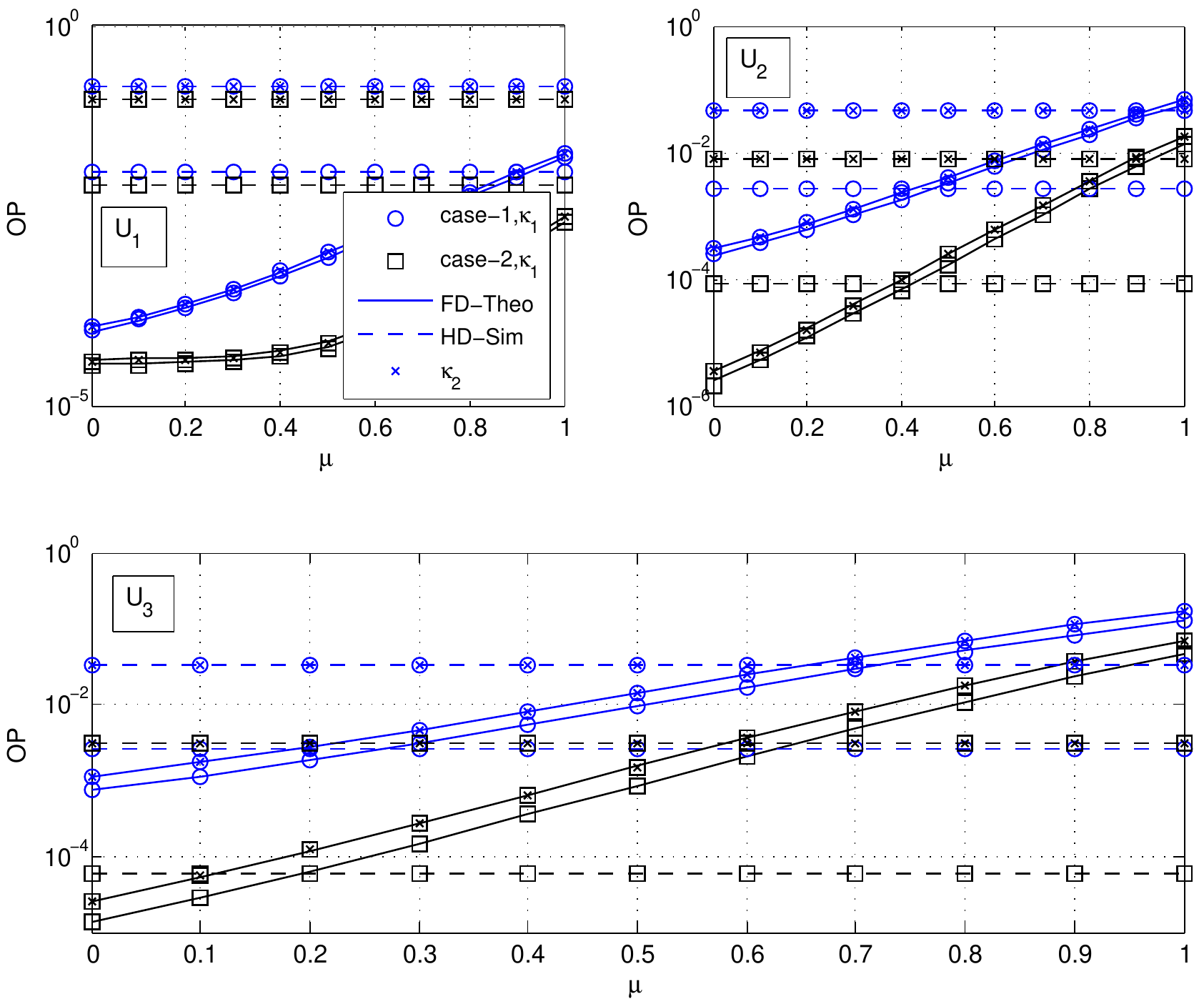}
	\caption{OP comparisons of the investigated FD-NOMA system with HD-NOMA counterpart versus $\mu$ in case fixed $SNR=15$ dB and different number of antenna and RHIs parameters.}
	\label{fig:7}
	\vspace{-2 mm}
\end{figure}

Fig. 7 represents OP comparisons of FD-NOMA (investigated) and HD-NOMA systems versus the quality of LI cancellation parameter $\mu$ in case fixed $SNR=15$ dB, $\sigma_{e,SR}^2=\sigma_{e,l}^2=0$, $\sigma_{ipsic}^2=0$, $m_{SR}=m_{LI}=m_{RU}=1$ and different RHIs parameters. In the figure, case-1:($N_S=2$;$N_D=2$), case-2:($N_S=3$;$N_D=2$), $\kappa_1$:($\kappa_{SR}=\kappa_{RU}=0$) and $\kappa_2$:($\kappa_{SR}=\kappa_{RU}=0.14$) definitions are made for simplicity. For fair comparison, threshold SIDNRs have the following relationship $\frac{1}{2}\log_2(1+\gamma_{th,l}^{HD})=\log_2(1+\gamma_{th,l}^{FD})$. Also, we set threshold SIDNRs of HD-NOMA as $\gamma_{th,1}^{HD}=0.9$, $\gamma_{th,2}^{HD}=1.5$ and $\gamma_{th,3}^{HD}=2$ to ensure satisfying the condition of $a_j$-$\gamma_{th,j}(\xi_j+\tilde{\xi}_j+\vartheta_1)>0$. As clearly seen from curves of the first user, FD-NOMA is better than HD-NOMA at the most of values $\mu$, however performance gap between them decreases as the quality of LI cancellation gets worse. On the other hand, according to the second and third users, FD-NOMA outperforms HD-NOMA when the value of $\mu$ is below $0.49$ and $0.26$ for the configuration of (case-1,$\kappa_1$), respectively. Moreover, in the presence of RHIs ($\kappa_2$), FD-NOMA is better than HD-NOMA when $\mu\leq0.9$ and $\mu\leq0.65$ for the second and third users. Consequently, the quality of LI cancellation under effect of RHIs can be worse than that of ideal HIs case for which FD-NOMA outperforms HD-NOMA.

Fig. 8 illustrates the OP performance of the investigated system versus RHIs parameters ($\kappa_{SR}=\kappa_{RU}$) for different effects of CEEs and ipSIC in case of fixed $SNR=15$ dB, $N_S=N_D=2$, $\mu=0.2$ and $m_{SR}=m_{LI}=m_{RU}=1$. For the figure to be better understandable, configurations are categorized into $5$ cases as following case-1:($\sigma_{e,SR}^2=0.03$; $\sigma_{ipsic}^2=0$; $\sigma_{e,l}^2=0.03$), case-2:($\sigma_{e,SR}^2=0.03$; $\sigma_{ipsic}^2=0$; $\sigma_{e,l}^2=0$), case-3:($\sigma_{e,SR}^2=0$; $\sigma_{ipsic}^2=0$; $\sigma_{e,l}^2=0.03$), case-4:($\sigma_{e,SR}^2=0$; $\sigma_{ipsic}^2=0.03$; $\sigma_{e,l}^2=0$) and case-5:($\sigma_{e,SR}^2=0$; $\sigma_{ipsic}^2=0$; $\sigma_{e,l}^2=0$). As clearly seen from the results of all users, OP performances get worse as the effect of RHIs increase, even OPs equal to $1$ under heavy RHIs effect. According to the first user, CEEs in the second hop worsen the performance more than that of in the first hop. However, in terms of the second and third users, CEEs in the first hop are much more effective than that of in the second hop. Moreover, ipSIC deteriorates the performance much more than CEEs in both hops for the second and third users.    

\begin{figure}[!t]
	\centering
	\includegraphics[width=0.76\columnwidth]{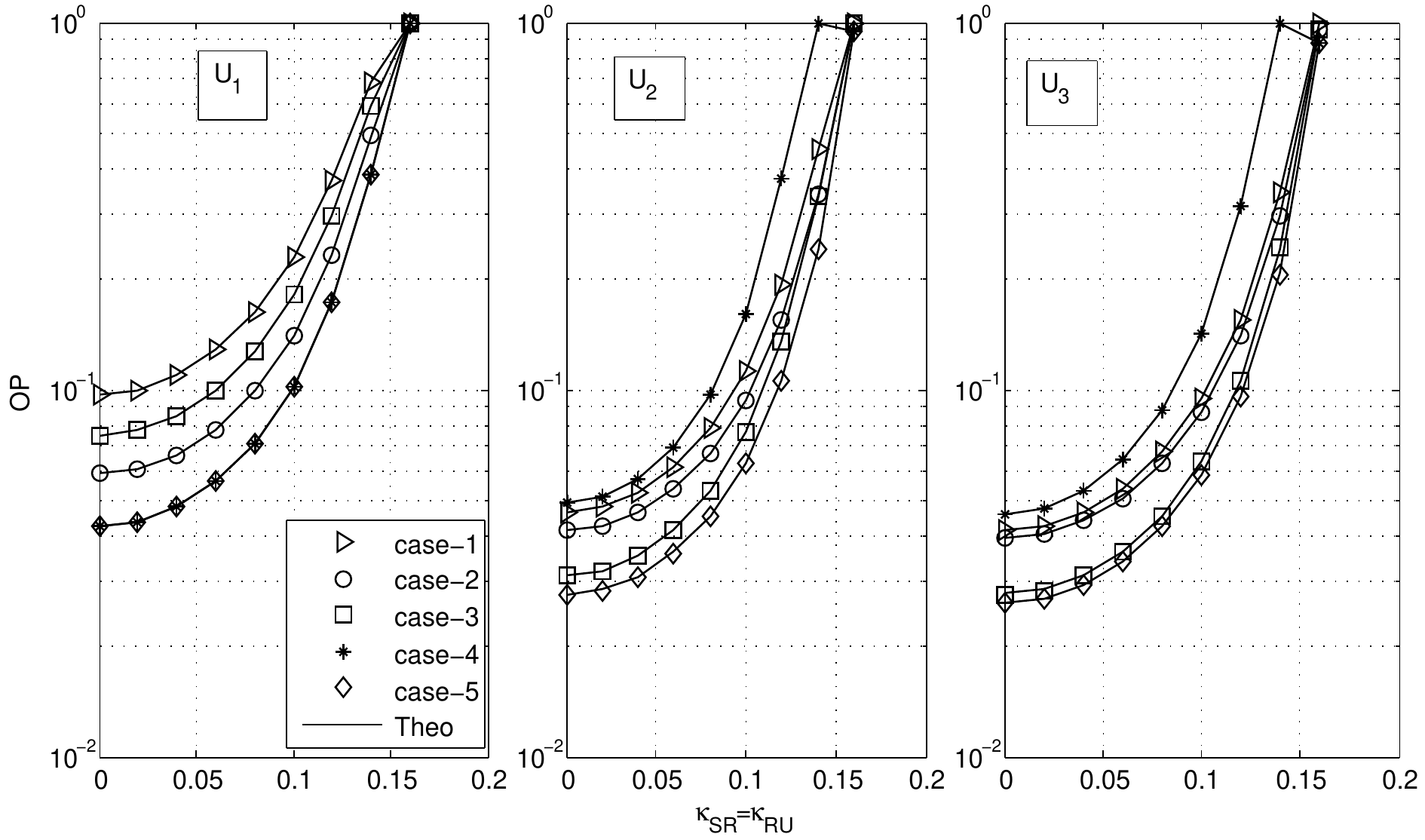}
	\caption{OP of users versus RHIs parameter $\kappa_{SR}=\kappa_{RU}$ in case of $\mu=0.2$, fixed $SNR=15$ dB and different CEEs and ipSIC configurations.}
	\label{fig:8}
	\vspace{-2 mm}
\end{figure}
\begin{figure}[!t]
	\centering
	\includegraphics[width=0.68\columnwidth]{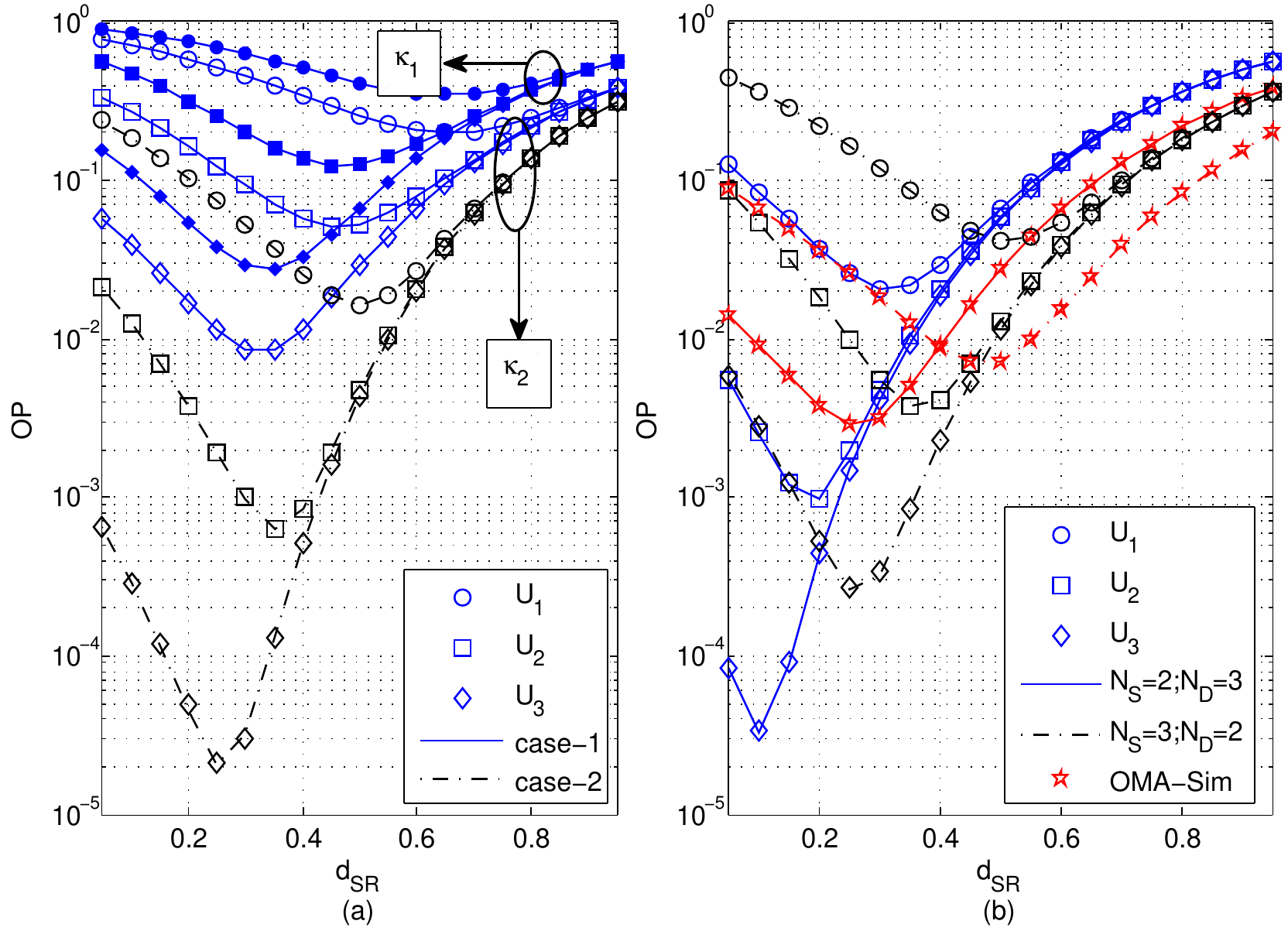}
	\caption{OP comparisons of the investigated FD-NOMA system with FD-OMA counterpart versus $d_{SR}$ in case of $\mu=0.2$, fixed $SNR=15$ dB, different fading and RHIs parameters, and number of antenna configurations.}
	\label{fig:9}
	\vspace{-2 mm}
\end{figure}

Fig. 9 depicts OP performance of the investigated system versus distances between the BS and relay ($d_{SR}$) in case of fixed $SNR=15$ dB and $\sigma_{e,SR}^2=\sigma_{e,l}^2=\sigma_{ipsic}^2=0$. The normalized distances in $R-U_l$ links are determined by $d_{RU_l}=1-d_{SR}$. In Fig. 9 (a), $\kappa_1$:($\kappa_{SR}=\kappa_{RU}=0.1$), $\kappa_2$:($\kappa_{SR}=\kappa_{RU}=0$), case-1:($N_S=2$;$N_D=1$, $m_{SR}=m_{LI}=m_{RU}=1$), case-2:($N_S=2$;$N_D=2$, $m_{SR}=2$;$m_{LI}=1$;$m_{RU}=1$). Also, Fig. 9 (b) is obtained for $\kappa_{SR}=\kappa_{RU}=0.1$, $m_{SR}=m_{LI}=m_{RU}=1$ and $\log_2(1+\gamma_{th}^{OMA})=\sum_{l=1}^{L}\log_2(1+\gamma_{th,l})$ relation is used to obtain OMA curves. From Fig. 9 (a), RHIs have the same level of effect on the performance at all values of distances for all users. It is observed that minimum OP can be achieved when $d_{SR}< d_{RU_l}$ for the second and third users while $d_{SR}\geq d_{RU_l}$ for the first user since optimum location of the relay has a close relation with diversity order and array gain which are also provided in asymptotic analyses. On the other hand, from Fig. 9 (b), FD-NOMA outperforms FD-OMA when the relay is close to the BS for the second and third users while FD-OMA is better at all values of $d_{SR}$ for the first user.

\section{Conclusion}
This paper analyzed the performance of MRT/MRC scheme in dual-hop NOMA FD AF relay networks over Nakagami-$m$ fading channels by considering the effects of RHIs. In addition, CEEs and ipSIC were also taken into account in order for the system be more realistic. For performance criterion, exact OP for any user was derived together with tight lower bound and asymptotic expressions. Numerical results demonstrated that performance of the investigated FD-NOMA system is strictly limited by error floor in the high SNR region, even all users are exposed to the same level, if LI cancellation can not be exploited. However, performance can be improved by increasing the number of antennas. On the other hand, all users can enjoy benefits of diversity order and array gain thanks to the quality of LI cancellation. Furthermore, our analysis revealed that the MRT beamforming is better than MRC on the performance improvement of users with lower power allocations than user with the highest power allocation. However, in case of CEEs, performance behavior trade-off between MRT and MRC schemes depends on imperfect channel conditions in both hops. Besides CEEs and imperfect LI cancellation, RHIs seriously deteriorate the performance, such that it is much more effective for the second and third users relative to the first user, while it has no effect on diversity order. On the other hand, under RHIs effect, LI cancellation process does not need to have very high quality when compared to ideal HIs case such that FD-NOMA outperforms HD-NOMA. It was also observed that imperfections which have the most and least deterioration effect on the performance are RHIs and CEEs, respectively. Furthermore, the minimum OP can be achieved with optimum relay location which has a close relation with diversity order and array gain, such that FD-NOMA outperforms FD-OMA when the relay is close to  the BS for users with lower power allocations.






\begin{thebibliography}{1}
	
\bibitem{QCLi}
Q. C. Li \textit{et al.}, \textquotedblleft 5G network capacity: key elements and technologies,\textquotedblright \emph{IEEE Veh. Technol. Mag.}, vol. 9, no. 1, pp. 71-78, Mar. 2014.
\label{bib:QCLi}

\bibitem{LiuY}
Y. Liu \textit{et al.}, \textquotedblleft Nonorthogonal multiple access for 5G and beyond,\textquotedblright~\emph{Proceedings of the IEEE}, vol. 105, no. 12, pp. 2347-2381, Dec. 2017.
\label{bib:LiuY}

\bibitem{Aldababsa}
M. Aldababsa \textit{et al.}, \textquotedblleft A tutorial on non-orthogonal multiple access for 5G and beyond,\textquotedblright~\emph{Wireless Commun. Mob. Comp.}, vol. 2018, Article ID 9713450, June 2018.
\label{bib:Aldababsa}

\bibitem{Saito1}
Y. Saito \textit{et al.}, \textquotedblleft Non-orthogonal multiple access (NOMA) for cellular future radio access,\textquotedblright~\emph{IEEE Veh. Tech. Conf.}, Dresden, Germany, Jun. 2013, pp. 1-5.
\label{bib:Saito1}

\bibitem{DingZ}
Z. Ding, Z. Yang, P. Fan and H. V. Poor, \textquotedblleft On the performance of non-orthogonal multiple access in 5G Systems with randomly deployed users,\textquotedblright~\emph{IEEE Signal Process. Lett.}, vol. 21, no. 12, pp. 1501-1505, Dec. 2014.
\label{bib:DingZ}

\bibitem{Timotheou}
S. Timotheou and I. Krikidis, \textquotedblleft Fairness for non-orthogonal multiple access in 5G systems,\textquotedblright~\emph{IEEE Signal Process. Lett.}, vol. 22, no. 10, pp. 1647-1651, Oct. 2015.
\label{bib:Timotheou}

\bibitem{Gui}
G. Gui, H. Sari and E. Biglieri, \textquotedblleft A new definition of fairness for non-orthogonal multiple access,\textquotedblright~\textit{IEEE Comm. Lett.}, vol. 23, no. 7, pp. 1267-1271, July 2019.
\label{bib:Gui}


\bibitem{DingZ1}
Z. Ding, F. Adachi, H.V. Poor, \textquotedblleft The application of MIMO to non-orthogonal multiple access,\textquotedblright~\emph{IEEE Trans. Wireless Commun.}, vol. 15, no. 1, pp. 537-552, Jan. 2016.
\label{bib:DingZ1}

\bibitem{QiSun}
Q. Sun, S. Han, C. I. and Z. Pan \textquotedblleft On the ergodic capacity of MIMO NOMA systems,\textquotedblright~\textit{IEEE Wireless Comm. Lett.}, vol. 4, no. 4, pp. 405-408, Aug. 2015.
\label{bib:QiSun}

\bibitem{MZeng}
M. Zeng \textit{et al.}, \textquotedblleft Capacity comparison between MIMO-NOMA and MIMO-OMA with multiple users in a cluster,\textquotedblright~\textit{IEEE J. Sel. Areas Commun.}, vol. 35, no. 10, pp. 2413-2424, Oct. 2017.
\label{bib:MZeng}

\bibitem{QZhang}
Q. Zhang, Q. Li and J. Qin, \textquoteleft Robust beamforming for nonorthogonal multiple-access systems in MISO channels,\textquoteright~\emph{IEEE Trans. Veh. Technol.}, vol. 65, no. 12, pp. 1231-1236, Dec. 2016.
\label{bib:QZhang}

\bibitem{DingZ2}
Z. Ding, R. Schober and H. V. Poor, \textquotedblleft A general MIMO framework for NOMA downlink and uplink transmission based on signal alignment,\textquotedblright~\emph{IEEE Trans. Wireless Commun.}, vol. 15, no. 6, pp. 4438-4454, June 2016.
\label{bib:DingZ2}

\bibitem{XChen}
X. Chen, Z. Zhang, C. Zhong and D. W. K. Ng, \textquotedblleft Exploiting multiple-antenna techniques for non-orthogonal multiple access,\textquotedblright~\textit{IEEE J. Sel. Areas Commun.}, vol. 35, no. 10, pp. 2207-2220, Oct. 2017.
\label{bib:XChen}

\bibitem{FAlavi}
F. Alavi, K. Cumanan, Z. Ding and A. G. Burr, \textquoteleft Beamforming techniques for nonorthogonal multiple access in 5G cellular networks,\textquoteright~\emph{IEEE Trans. Veh. Technol.}, vol. 67, no. 10, pp. 9474-9487, Oct. 2018.
\label{bib:FAlavi}

\bibitem{MtokaCL}
M. Toka and O. Kucur, \textquotedblleft Non-orthogonal multiple access with Alamouti space–time block coding,\textquotedblright~\emph{IEEE Commun. Lett.}, vol. 22, no. 9, pp. 1954-1957, Sept. 2018.
\label{bib:MtokaCL}

\bibitem{Tarokh}
V. Tarokh, H. Jafakhani, and A. R. Calderbank, \textquotedblleft Space-time block codes from orthogonal designs,\textquotedblright~\textit{IEEE Trans. Inf. Theory}, vol. 45, no. 5, pp. 1456-1467, Jul. 1999.
\label{bib:Tarokh}

\bibitem{Mtoka}
M. Toka and O. Kucur, \textquotedblleft Performance analysis of OSTBC-NOMA system in the presence of practical impairments,\textquotedblright~\emph{IEEE Trans. Veh. Technol.}, vol. 69, no. 9, pp. 9697-9706, Sept. 2020.
\label{bib:Mtoka}

\bibitem{YYu1}
Y. Yu, H. Chen, Y. Li, Z. Ding, L. Song and B. Vucetic, \textquotedblleft Antenna selection for MIMO nonorthogonal multiple access systems,\textquotedblright~\textit{IEEE Trans. Veh. Technol.}, vol. 67, no. 4, pp. 3158-3171, Apr. 2018.
\label{bib:YYu1}

\bibitem{AldababsaMajTAS}
M. Aldababsa, and O. Kucur, \textquotedblleft Majority based antenna selection schemes in downlink NOMA network with channel estimation errors and feedback delay,\textquotedblright~\emph{IET Commun.}, vol. 14, no. 17, pp. 2931-2943, Oct. 2020.
\label{bib:AldababsaMajTAS}


\bibitem{DingZ3}
Z. Ding, M. Peng and H. V. Poor \textquotedblleft Cooperative non-orthogonal multiple access in 5G Systems,\textquotedblright~\emph{IEEE Commun. Lett.}, vol. 19, no. 8, pp. 1462-1465, Aug. 2015.
\label{bib:DingZ3}

\bibitem{JBKim}
J. B. Kim, and I. H. Lee, \textquotedblleft Non-orthogonal multiple access in coordinated direct and relay transmission,\textquotedblright~\emph{IEEE Commun. Lett.}, vol. 19, no. 11, pp. 2037-2040, Nov. 2015.
\label{JBKim}

\bibitem{JJmen}
J. Men, J. Ge, and C. Zhang, \textquotedblleft Performance analysis for downlink relaying aided non-orthogonal multiple access networks with imperfect CSI over Nakagami-${m}$ Fading,\textquotedblright~\textit{IEEE Access}, vol. 5, pp. 998-1004, Feb. 2017.
\label{bib:JJmen}

\bibitem{ZhangY}
Y. Zhang, J. Ge, and E. Serpedin, \textquotedblleft Performance analysis of non-orthogonal multiple access for downlink networks with antenna selection over Nakagami-$m$ fading channels,\textquotedblright~\emph{IEEE Trans. Veh. Technol.}, vol. 66, no. 11, pp. 10590-10594, Nov. 2017.
\label{bib:ZhangY}

\bibitem{XYan}
X. Yan, J. Ge, Y. Zhang and L. Gou, \textquotedblleft NOMA-based multiple-antenna and multiple-relay networks over Nakagami-m fading channels with imperfect CSI and SIC error,\textquotedblright~\emph{IET Commun.}, vol. 12, no. 17, pp. 2087-2098, Oct. 2018.
\label{bib:XYan}

\bibitem{HLi}
H. Li, J. Li and L. Lv, \textquotedblleft Joint relay-and-antenna selection for cooperative non-orthogonal multiple access,\textquotedblright~\textit{IET Commun.}, vol. 13, no. 13, pp. 2012-2019, Aug. 2019.
\label{bib:HLi}

\bibitem{AldababsaMRTRAS}
M. Aldababsa and O. Kucur, \textquotedblleft Performance of cooperative multiple-input multiple-output NOMA in Nakagami-m fading channels with channel estimation errors,\textquotedblright~\emph{IET Commun.},~ vol. 14, no. 2,~ pp. 274-281, Jan. 2020.
\label{bib:AldababsaMRTRAS}

\bibitem{Duarte}
M. Duarte, C. Dick and A. Sabharwal, \textquoteleft Experiment-driven characterization of full-duplex wireless systems,\textquoteright~\emph{IEEE Trans. Wireless Commun.}, vol. 11, no. 12, pp. 4296-4307, Dec. 2012.
\label{bib:Duarte}

\bibitem{Rodriguez}
L. J. Rodriguez, N. H. Tran, T. Le-Ngoc, \textquoteleft Performance of full duplex af relaying in the presence of residual self-interference,\textquoteright~\emph{IEEE J. Sel. Areas Commun.}, vol. 32, no. 9, pp. 1752-1764, Sept. 2014.
\label{bib:Rodriguez}


\bibitem{ZhongC}
C. Zhong, and Z. Zhang, \textquotedblleft Non-orthogonal multiple access with cooperative full-duplex relaying,\textquotedblright~\emph{IEEE Commun. Lett.}, vol. 20, no. 12, pp. 2478-2481, Dec. 2016.
\label{ZhongC}

\bibitem{TMCC}
T. M. C. Chun and H.-J. Zepernick, \textquotedblleft Performance of non-orthogonal multiple access system with full-duplex relaying,\textquotedblright~\textit{IEEE Commun. Lett.}, vol. 22, no. 10, pp. 2084-2087, Oct. 2018.
\label{bib:TMCC}

\bibitem{YAlsaba}
Y. Alsaba, C. Y. Leow and S. K. Abdul Rahim, \textquotedblleft Full-duplex cooperative non-orthogonal multiple access with beamforming and energy harvesting,\textquotedblright~\textit{IEEE Access}, vol. 6, pp. 19726-19738, Apr. 2018.
\label{bib:YAlsaba}

\bibitem{Mohammadi}
M. Mohammadi, B. K. Chalise, A. Hakimi, Z. Mobini, H. A. Suraweera and Z. Ding, \textquotedblleft Beamforming design and power allocation for full-duplex non-orthogonal multiple access cognitive relaying,\textquotedblright~\emph{IEEE Trans. Commun.}, vol. 66, no. 12, pp. 5952-5965, Dec. 2018
\label{bib:Mohammadi}


\bibitem{Schenk}
T. Schenk, \textit{RF Imperfections in High-Rate Wireless Systems: Impact and Digital Compensation}, Dordrecht, The Nederlands: Springer, 2008. 
\label{bib:Schenk}

\bibitem{Bjornson}
E. Bjornson, M. Matthaiou, and M. Debbah, \textquotedblleft A new look at dual-hop relaying: performance limits and hardware impairments,\textquotedblright~\emph{IEEE Trans. Commun.}, vol. 61, no. 11, pp. 4512-4525, Nov. 2013.
\label{bib:Bjornson}

\bibitem{Bjornson2}
E. Bjornson, J. Hoydis, M. Kountouris and M. Debbah, \textquotedblleft Massive MIMO systems with non-ideal hardware: energy efficiency, estimation, and capacity limits,\textquotedblright~\emph{IEEE Trans. Inf. Theory}, vol. 60, no. 11, pp. 7112-7139, Nov. 2014.
\label{bib:Bjornson2}

\bibitem{XLi3}
X. Li \textit{et al.}, \textquotedblleft Security analysis of multi-antenna NOMA networks under I/Q imbalance,\textquotedblright~\emph{Electronics}, vol. 8, 1327, Nov. 2019.
\label{bib:XLi3}

\bibitem{MtokaMRTRAS}
M. Toka, and O. Kucur, \textquotedblleft Performance of MRT/RAS MIMO-NOMA with residual hardware impairments,\textquotedblright~\textit{IEEE Wireless Commun. Lett.}, Early Access, Feb. 2021.
\label{bib:MtokaMRTRAS}

\bibitem{FDing}
F. Ding \textit{et al.}, \textquotedblleft Impact of residual hardware impairments on non-orthogonal multiple access based amplify-and-forward relaying networks,\textquotedblright~\emph{IEEE Access}, vol. 6, pp. 15117-15131, Mar. 2018.
\label{bib:FDing}

\bibitem{XLi}
X. Li \textit{et al.}, \textquotedblleft Residual transceiver hardware impairments on cooperative NOMA networks,\textquotedblright~\emph{IEEE Trans. Wireless Commun.}, vol. 19, no. 1, pp. 680-695, Jan. 2020.
\label{bib:XLi}

\bibitem{XLi2}
X. Li \textit{et al.}, \textquotedblleft Joint effects of residual hardware impairments and channel estimation errors on SWIPT assisted cooperative NOMA networks,\textquotedblright~\emph{IEEE Access}, vol. 7, pp. 135499-135513, 2019.
\label{bib:XLi2}

\bibitem{CBLe}
C. B. Le, D. T. Do and M. Voznak, \textquotedblleft Exploiting impact of hardware impairments in NOMA: adaptive transmission mode in FD/HD and application in internet-of-things,\textquotedblright~\emph{Sensors}, vol. 19, 1293, Mar. 2019.
\label{bib:CBLe}

\bibitem{CDeng}
C. Deng, M. Liu, X. Li and Y. Liu, \textquotedblleft Hardware impairments aware full-duplex NOMA networks over rician fading channels,\textquotedblright~\emph{IEEE Access}, vol. 7, pp. 135499-135513, 2019.
\label{bib:CDeng}

\bibitem{LoT}
T. Lo, \textquotedblleft Maximal-ratio transmission,\textquotedblright~\emph{IEEE Trans. Commun.}, vol. 47, no. 10, pp. 1458-1461, Oct. 1999.
\label{bib:LoT}

\bibitem{Simon}
M. Simon, M.-S. Alouini, \textit{Digital communication over fading channels}, 2nd ed. London: Wiley, 2005.
\label{bib:Simon}

\bibitem{Medard}
M. Medard, \textquotedblleft The effect upon channel capacity in wireless communications of perfect and imperfect knowledge of the channel,\textquotedblright~\emph{IEEE Trans. Inf. Theory}, vol. 46, no. 3, pp. 933-946, May 2000.
\label{bib:Medard}

\bibitem{Nguyen}
B. C. Nguyen \textit{et al.}, \textquoteleft Impact of hardware impairments on the outage probability and ergodic capacity of one-way and two-way full-duplex relaying systems,\textquoteright~\emph{IEEE Trans. Veh. Technol.}, vol. 69, no. 8, pp. 8555-8567, Aug. 2020.
\label{bib:Nguyen}


\bibitem{HasanA}
A. Hasan, and J. Andrews, \textquotedblleft Cancellation error statistics in a power-controlled CDMA system using successive interference cancellation,\textquotedblright~in Proc. \emph{IEEE Int. Symp. Spread Spect. Techn. App.}, Sydney, NSW, Australia, Sept. 2004.
\label{bib:HasanA}

\bibitem{Tweed}
D. Tweed \textit{et al.}, \textquotedblleft Outage-constrained resource allocation in uplink NOMA for critical applications,\textquotedblright~\emph{IEEE Access}, vol. 5, pp. 27636-27648, Dec. 2017.
\label{bib:Tweed}

\bibitem{Gradshteyn}
I. S. Gradshteyn, and I. M. Ryzhic, \textit{Table of Integrals, series, and Products}, 6th ed.~~New York, NY, USA: Academic, 2000. 
\label{bib:Gradshteyn}

\bibitem{DavidHA}
H. A. David and H. N. Nagaraja, \textit{Order Statistics}, 3rd ed.~~Hoboken, NJ, USA: Wiley, 2003. 
\label{bib:DavidHA}

\bibitem{ZWang}
Z. Wang, and G. B. Giannakis, \textquotedblleft A simple and general parameterization quantifying performance in fading channels,\textquotedblright~\textit{IEEE Trans. Commun.}, vol. 51, no. 8, pp. 1389-1398, Aug. 2003.
\label{bib:ZWang} 

\bibitem{Oldham}
K. B. Oldham, J. Myland, and J. Spanier, \textit{An Atlas of Functions with Equator the Atlas Function Calculator}, 2nd ed.~~Springer, 2008. 
\label{bib:Oldham}




\end{thebibliography}
\end{document}